\documentclass{emulateapj-rtx4}
\usepackage{amssymb, amsmath, epsfig}

\newcommand{\comment}{}

\newcommand{\HI}{H{\sc ~i}}

\newcommand{\HeII}{He{\sc ~ii}}

\begin{document}

\title{the case against large intensity fluctuations in the $z\sim 2.5$ \HeII\ Ly$\alpha$ forest}
\author{Matthew McQuinn\altaffilmark{1}, Gabor Worseck\altaffilmark{2}}

\altaffiltext{1} {Department of Astronomy, University of California, Berkeley, CA 94720, USA; mmcquinn@berkeley.edu}
\altaffiltext{2}{Max Planck Institute for Astronomy, K\"{o}nigstuhl 17, D-69117 Heidelberg, Germany\\}

\begin{abstract} 
Previous studies of the $2.2 < z< 2.7$ \HeII\ Ly$\alpha$ forest measured much larger  ionizing background fluctuations than are anticipated theoretically.   We re--analyze recent Hubble Space Telescope data from the two \HeII\ sightlines that have been used to make these measurements, HE2347-4342 and HS1700+6416, and find that the vast majority of the absorption is actually consistent with a single, spatially uniform \HeII\ photoionization rate.  We show that the data constrain the RMS fractional fluctuation level smoothed over $1~$Mpc to be $< 2$ and discuss why other studies had found $10$ times larger fluctuations.  Our measurement is consistent with models in which quasars dominate the $z=2.5$ metagalactic \HeII--ionizing background (but it can accommodate less compelling models), and it suggests that quasars (rather than stars) are the dominant contributor to the HI--ionizing background.  We detect a \HeII\ transverse proximity effect that is slightly offset in redshift from a known quasar.  Its profile and offset may indicate that the quasar turned on $10~$Myr ago. 
\end{abstract}

\keywords{diffuse radiation --- intergalactic medium --- quasars: absorption lines}

\section{introduction}

During reionization processes, the hydrogen and/or helium ionizing backgrounds were expected to fluctuate wildly \citep{miralda00, furlanettoJfluc, furlanettodixon, mesinger09, mcquinn-LL}.  Afterwards, these fluctuations should have decreased sizably, as the mean free path of ionizing photons became much larger than the mean inter-source separation.  A uniform hydrogen--ionizing background is assumed in the cosmological simulations that are used to constrain cosmological parameters from the $2 \lesssim z\lesssim 4$ Ly$\alpha$ forest (e.g., \citealt{croft02, viel04, mcdonald05b}).  The consistency of these cosmological parameter estimates with the concordance cosmology supports this assumption, but it has yet to be directly validated.    

The most direct measure of fluctuations in the ultraviolet background intensity field derives from the comparison of absorption in the \HeII\ Ly$\alpha$ forest to coeval absorption in the \HI\ Ly$\alpha$ forest.  In particular, the ratio of the coeval optical depths in these two lines provides an estimate for $\eta$ -- a quantity proportional to the ratio of the $1~$Ry photon density to the $4~$Ry density and that we will define shortly \citep{miralda90}.  
Only two existing \HeII\ Ly$\alpha$ forest observations, towards quasars HE2347-4342 and HS1700+6416, offer sufficient signal-to-noise ratios ($S/N$) to place interesting constraints on $\eta$ at $z < 2.7$ -- redshifts after \HeII\ reionization appears to have completed \citep{shull10}.  In both of these spectra, previous studies concluded that large fluctuations in $\eta$ are present \citep{shull04, zheng04, fechner06, fechner07, shull10, syphers13}.  \citet{shull04} also noted strong correlations between these fluctuations and the local density (using \HI\ absorption as its proxy) on many tens down to $\sim 1~$proper Mpc scales (but see \citealt{fechner07}).
  
\citet{bolton06} argued that the level of $\eta$ fluctuations (and their correlations with density) simply owes to the sources of $4~$Ry photons (i.e., quasars) being rare and to the post--\HeII\ reionization mean free path of \HeII--ionizing photons being short.   The models in \citet{bolton06} assumed a mean free path of $40~$comoving Mpc for $4~$Ry photons at $z= 2.5$, comparable to the space density of their sources -- $L_*$ quasars --, a situation that leads to large fluctuations in the $4~$Ry background.  However, recent theoretical estimates for the $4~$Ry mean free path, $\lambda_{\rm HeII}$, find values that are much larger than assumed in \citet{bolton06}, $\lambda_{\rm HeII} = 100-200~$comoving Mpc \citep{faucher09, davies12}.  The newer mean free path estimates reduce the expected amplitude of $\eta$ fluctuations and, hence, may make the \citet{bolton06} explanation untenable.

It is timely to revisit the constraints on and the interpretation of $\eta$ as HE2347-4342 and HS1700+6416 have recently been re-observed using the Cosmic Origins Spectrograph (COS) on the Hubble Space Telescope.  (See \citealt{shull10} and \citealt{syphers13} for a comparison with the previous Far Ultraviolet-Visual Echelle Spectrograph [FUSE] data.)  While the $S/N$ of the COS and FUSE data when binned at the same resolution are comparable, background subtraction can be done more reliably with COS.  The recent analysis of HE2347-4342 by \citet{shull10} again found $\sim 1$~dex fluctuations in $\eta$, claiming consistency with the previous findings in \citet{shull04} using FUSE.  Similarly, \citet{syphers13} found large $\eta$ fluctuations towards HS1700+6416.  We re-analyze the COS HE2347-4342 and HS1700+6416 data here, using improved methods to fit the continuum and estimate $\eta$, and instead find (1) no compelling evidence for large $\eta$ fluctuations and (2) that the data can be explained with standard ultraviolet background models.

 This paper is organized as follows.  Section \ref{sec:HE2347} reanalyzes the $z \lesssim 2.7$ HE2347-4342 data.  We provide new estimates for $\eta(z)$, as well as a detailed investigation of the most important systematic -- continuum fitting of the \HI\ data.
Section \ref{sec:model} compares the observations to sophisticated models for $\eta$ fluctuations.  We find the level of fluctuations in the ionizing backgrounds inferred from our $\eta$ estimates to be consistent with the predictions of standard ionizing background models.  Section~\ref{sec:implications} discusses implications of our analysis for (1) quasar lifetimes and (2) the quasar and stellar contributions to the \HI--  and \HeII--ionizing backgrounds.  Appendix A  applies our analysis pipeline to HS1700+6416, which yields weaker (albeit consistent) constraints on $\eta$.  The weaker constraints for this sightline owe to apparent structure in its optical continuum, which contributes uncertainty to the continuum placement.

Throughout, we (re)define $\eta$ as the ratio of \HeII\ to \HI\ number density (as opposed to its historical definition as the column density ratio of these species) such that
  \begin{equation}
  \eta \equiv \frac{n_{\rm HeII}}{n_{\rm HI}} \approx 4 \frac{\tau_{\rm HeII}}{\tau_{\rm HI}} \approx 0.43 \frac{\Gamma_{\rm HI}}{\Gamma_{\rm HeII}},
  \label{eqn:etarels}
  \end{equation}
  where $\tau_X$ and $\Gamma_X$ are respectively the  optical depth and photoionization rate for ionic species $X$.  The first approximate relation in Eq.~(\ref{eqn:etarels}) applies in the so-called limit of pure turbulent broadening (i.e., negligible thermal broadening), which we will show is a good approximation, and the latter relation also assumes photoionization equilibrium and ignores a weak temperature dependence.  
We use hats to denote estimated quantities throughout.  For example, $\widehat{\eta}(z)$ is the estimated value of $\eta$ at redshift $z$.  All power-law indices are defined as being \emph{minus} the logarithmic slope.  Our calculations assume a flat $\Lambda$CDM cosmology with $Y_{\rm He} = 0.24$, $h=0.7$, $\Omega_m = 0.27$, $\Omega_b= 0.046$, $\sigma_8 = 0.8$, and $n_s = 0.96$ \citep{larson11}.

\section{$\eta$ fluctuations in the \HeII\ Ly$\alpha$ forest}
\label{sec:HE2347} 

\subsection{data}
\label{sec:data}

\begin{figure}
\begin{center}
{\epsfig{file=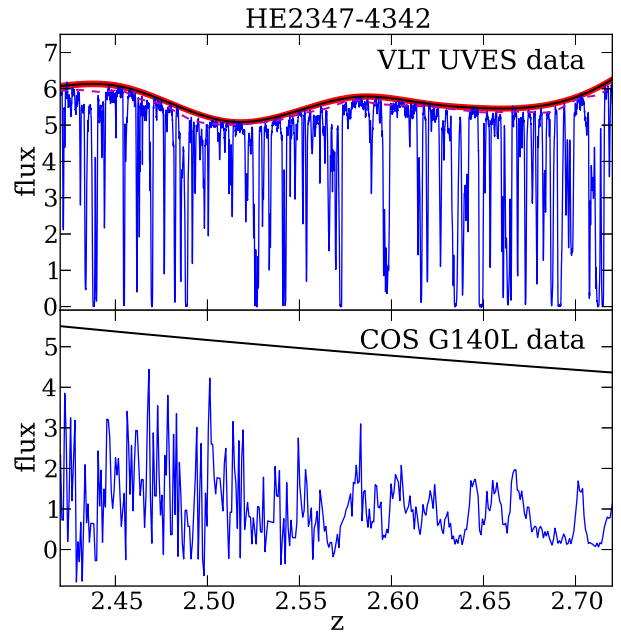, width=8.9cm}}
\end{center}
\caption{The top panel shows the UVES \HI\ Ly$\alpha$ forest spectrum of HE2347-4342 at $2.43 < z< 2.72$ as well as our continuum fits.  The blue solid curve is the UVES measurement, the black solid line shows the continuum estimate from our continuum--fitting algorithm (cf. Section~\ref{sec:data}), the red band shows the error in the continuum estimate (calculated by applying the fitting algorithm to the mocks), and the magenta dashed curve is the previous continuum fit of \citet[which is also taken as our mocks' continua]{worseck07}.  The bottom panel shows the COS G140L \HeII\ Ly$\alpha$ forest spectrum of HE2347-4342 (blue curve) and the continuum estimate from \citet[black curve]{worseck13}.  The flux units are arbitrary in both panels.
\label{fig:contfits}}
\end{figure}


HE2347-4342 is the brightest known and most studied \HeII\ Ly$\alpha$ forest sightline, having a slightly higher flux in the extreme ultraviolet than the other sightline we analyze, HS1700+6416.  Numerous studies have analyzed the \HeII\ absorption towards HE2347-4342 \citep{reimers97, kriss01, smette02, shull04, zheng04, fechner07, worseck07, shull10, muzahid11}.  
We use observations of this sightline with the COS G140L grating, which covers the \HeII\ Ly$\alpha$ forest spectral region, and with the Very Large Telescope (VLT) Ultraviolet-Visual Echelle Spectrograph (UVES) for the \HI\ Ly$\alpha$ forest (see Fig.~\ref{fig:contfits}).  See \citet{worseck13} for the details of how these spectra were processed.\footnote{Of note, we calibrated the wavelengths of the G140L data by matching to features in the FUSE data, as was also done in \citet{syphers13}.  We do not think that wavelength calibration is a major uncertainty in our analysis.}  UVES captures the \HI\ Ly$\alpha$ forest with $S/N = 100$ at a resolution of $\lambda/\Delta \lambda = 45,000$, which is sufficient to resolve all the \HI\ absorption features.  COS G140L has $\lambda/\Delta \lambda \approx 1800$ at the wavelengths of interest and, hence, does not resolve most lines in the \HeII\ Ly$\alpha$ forest.  The low resolution of the COS G140L grating allows us to measure $\eta$ smoothed over $\approx 2~$comoving Mpc scales.  Fortunately, the bulk of the fluctuations in $\eta$ are anticipated to be coherent over larger scales (Section \ref{sec:model}).

The \HeII\ Ly$\alpha$ forest continuum of HE2347-4342 is estimated by extrapolating a power-law fit to the continuum redward of the \HeII\ forest as described in \citet{worseck11}.  The black curve in the bottom panel of Figure~\ref{fig:contfits} shows the estimated HE2347-4342 continuum.\footnote{For HS1700 (Appendix A), partial Lyman-limit absorption is also corrected for when estimating the continuum, but there are no detected partials in the HE2347 spectrum.}  {\comment A power-law will miss features in the continuum and we estimate err at the $\lesssim10\%$ level.  Such errors lead in turn to $\lesssim10\%$ errors in $\widehat{\eta}$, much smaller than other sources of uncertainty.}   However, uncertainty in the \HI\ Ly$\alpha$ forest continuum estimate leads to larger errors in $\widehat{\eta}$, as voids in the \HI\ Ly$\alpha$ forest mistakenly placed at the continuum would yield infinite $\widehat{\eta}$.  Much of the transmission in the \HeII\ forest in fact occurs in void regions (as we will show that the optical depth in \HeII\ is $\sim 25 \times$ larger than the optical depth in \HI).

To quantify the impact of errors in the \HI\ continuum estimate on $\widehat{\eta}$, we generated mock \HI\ spectra by stitching together \HI\ Ly$\alpha$ forest skewers calculated from our fiducial cosmological simulation (discussed shortly) with the mean absorption in the \HI\ Ly$\alpha$ forest normalized to the measurement of \citet{faucher07}.  Redshift evolution is included by using multiple temporal snapshots from the simulations (which were output every $\Delta z = 0.05$), and the mocks have identical wavelength resolution and $S/N$ values as the UVES data.  Furthermore, the HE2347-4342 estimated HI Ly$\alpha$ continuum from \citet{worseck07} is used as the mocks' continua.  

We applied an automated continuum fitter to the data as well as the mocks.  In particular, this fitter uses the following algorithm with the true flux as the initial input for the `estimated normalized flux':
\begin{enumerate}
\item Average estimated normalized flux over $5$ nearest pixels to reduce noise (equivalent to $17$~km~s$^{-1}$).
\item Locate pixels with the largest estimated normalized flux every $3500~$km~s$^{-1}$ ($\Delta z \approx 0.04$).
\item A cubic spline fit that intersects $0.005$ above each of these pixels is the new estimated continuum.
\item Divide the observed flux by the estimated continuum to generate the new normalized flux estimate.
\item Repeat the above steps on the new normalized flux, terminating when the continuum estimate has converged.
\end{enumerate}
This algorithm was selected based on its performance on the mocks, and it was finalized prior to applying it to measure $\eta$ from the HE2347 data. 
The value of $0.005$ in step $2$ is approximately the smallest optical depth in the Ly$\alpha$ forest mocks (occurring every $\sim 10^4~$km~s$^{-1}$).  
We find that this continuum fitting algorithm results in a bias in the mocks' continuum estimate of $+0.4\%$ and a root mean square (RMS) continuum error of $1.2\%$.  {\comment Because the RMS error is significantly larger than $0.005$, our results are \emph{not} sensitive to this somewhat arbitrary choice.}  The red band in Figure~\ref{fig:contfits} shows this error centered around this algorithm's estimated continuum (black solid curve) and also the measured UVES spectrum (blue solid curve).   This algorithm should not be applied more generally to fit \HI\ Ly$\alpha$ forest data as it is optimized for HE2347, which not only affords high $S/N ~(\sim 100)$ but also has relatively little continuum structure.  

\begin{figure*}
\begin{center}
{\epsfig{file=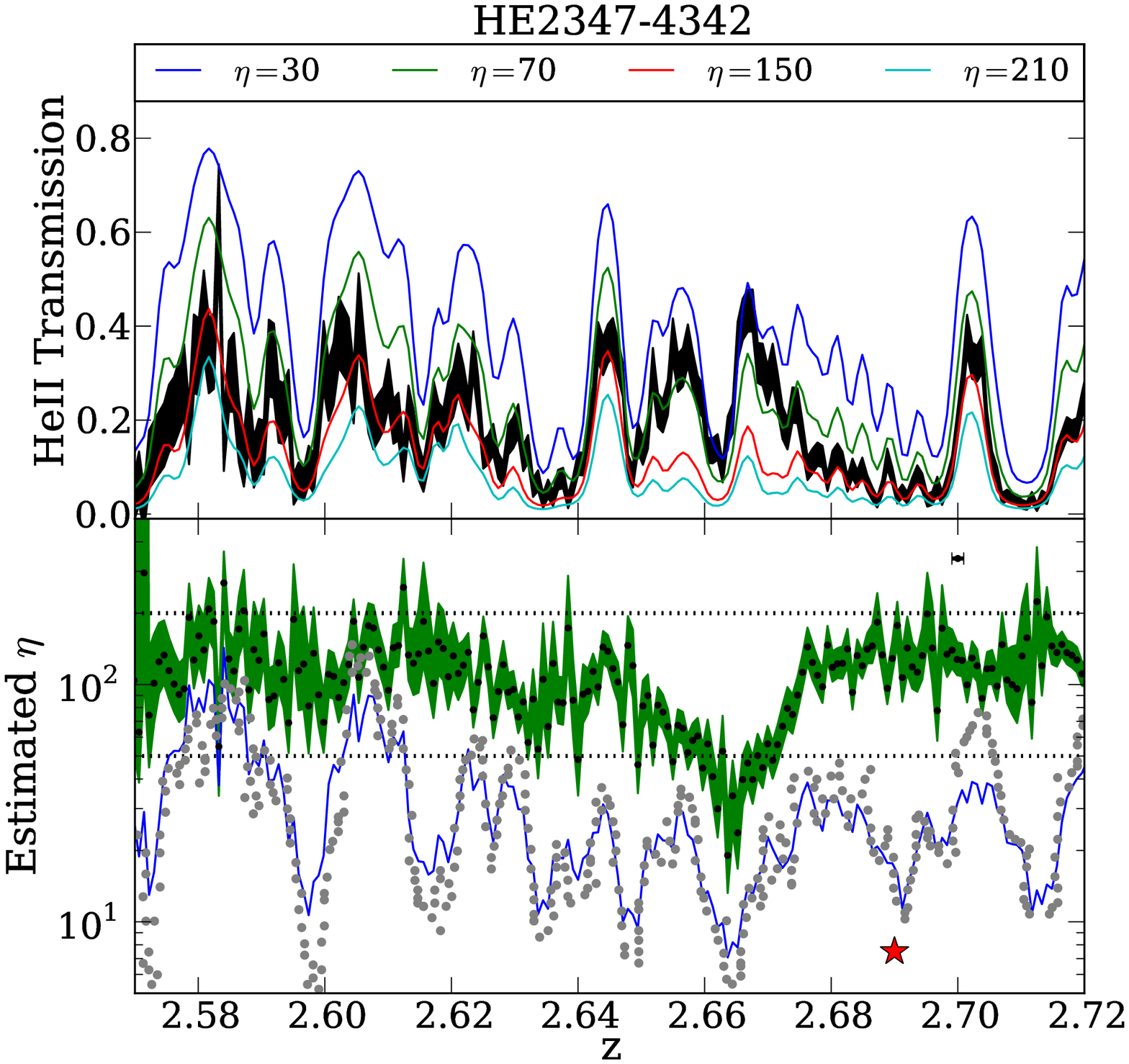, width=8.8cm}}
{\epsfig{file=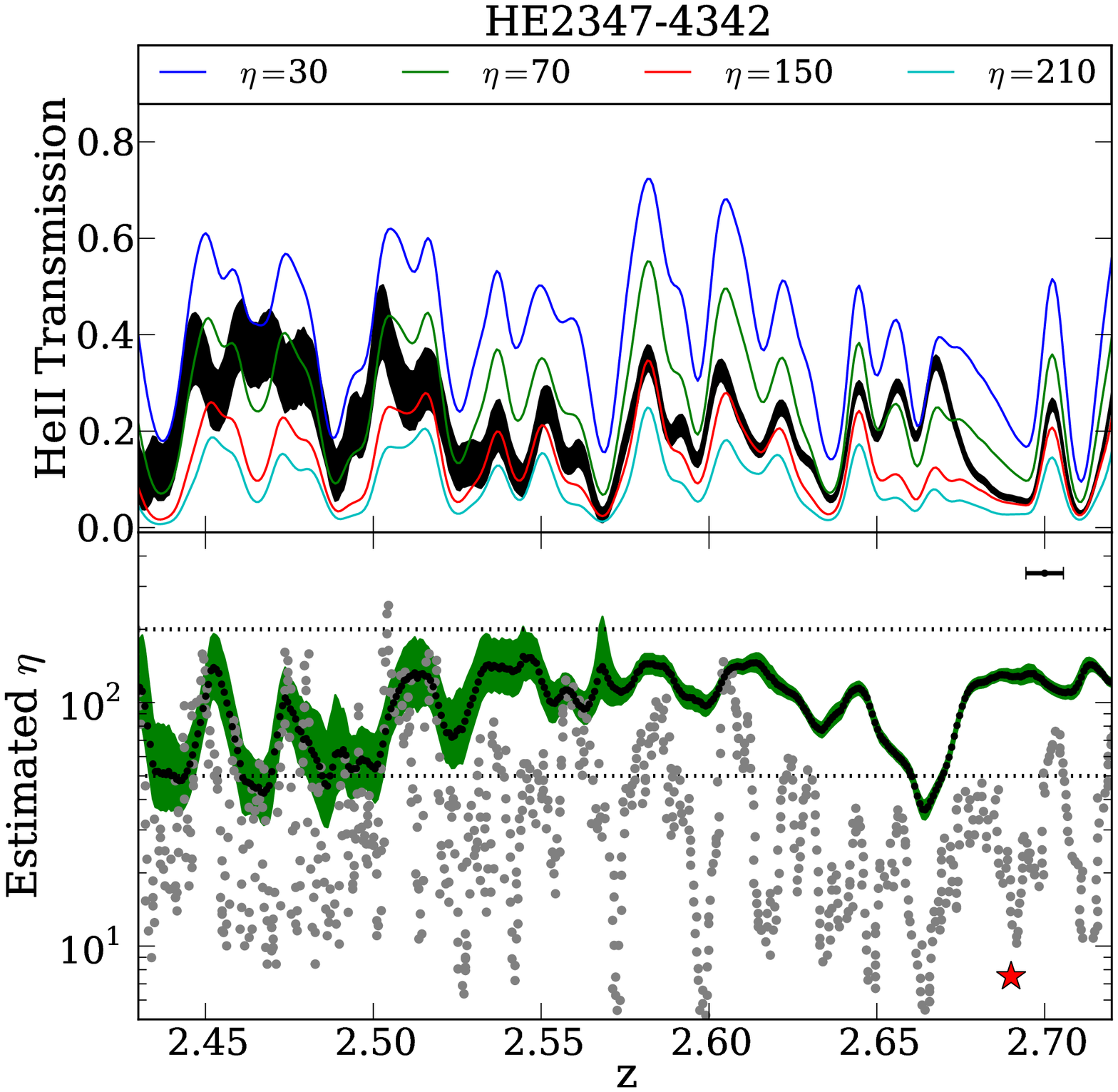, width=8.8cm}}
\end{center}
\caption{$\eta$ estimates towards HE2347-4342 at $2.43 < z < 2.72$.  The colored thin solid curves in the top panels show the estimated \HeII\ transmission for different values of $\eta$, calculated by forward--modeling the \HI\ Ly$\alpha$ transmission and then convolving with the COS G140L LSF with FWHM~$7~$COS pixels (left panel) or this plus a Gaussian with FWHM of $14$~COS pixels (right panel).  The black band in the top panels is the measured \HeII\ transmission with COS G140L grating, where the width shows the $1~ \sigma$ statistical errors.  The black points in the bottom panel are the $\widehat{\eta}$ inferred from fitting the forward--modeled \HI\ absorption.  The green regions show the error on $\widehat{\eta}$, and the grey points are the previous $\eta$ estimates using the same COS G140L spectrum \citep{shull10}.  The blue thin solid curve in the bottom--left panel (which roughly traces the \citealt{shull10} points) is $\eta$ estimated from data put through our pipeline but using the \citet{shull10} estimator, $\widehat{\eta}_{\rm simple}$.  All curves and the black dotted points are sampled by binning at a resolution of $2$ COS pixels.  The horizontal dotted lines show $\eta = 50$ and $200$ for reference.  The red star in the bottom panels is the redshift of the only identified proximate quasar within a transverse separation of $40$~comoving Mpc, which likely sources the $z = 2.66$ feature in our $\widehat{\eta}$.
\label{fig:data}}
\end{figure*} 

The continuum placement has some dependence on the intergalactic thermal history as hotter regions have less absorption and, thus, are more likely to fall closer to the continuum \citep{lee12}.  To address this issue (as well as to study other biases of our analysis pipeline with mock spectra), we ran three $2\times 512^3$ particle, $25/h~$comoving Mpc smooth particle hydrodynamics simulations using the Gadget-3 code \citep{springel05} initialized with densities and fluctuations consistent with the specified cosmology.  The convergence tests in Appendix C of \citet{lidz10} suggest that these box size and particle numbers are sufficient to resolve the low-density $z=3$ IGM.\footnote{One might worry that we are not resolving structures on the scale of the splined points with these simulations.  However, the large-scale 1D Ly$\alpha$ forest power spectrum is very white \citep{mcquinn-Tfluc, mcquinnwhite}, which indicates that the Poissonian fluctuations from small-scale absorbers (that are resolved in these simulations) dominates over their missing large-scale correlations.}   The first of these simulations assumes optically thin photoheating using the \citet{faucher09} ultraviolet background model.  Optically thin heating inevitably results in temperatures that are too low \citep{abel99}.  The second -- which we take as our fiducial simulation -- is the same as the first except that it doubles the temperature of all gas particles at $z=3.5$ to emulate the expected heating from \HeII\ reionization \citep{mcquinn-HeII}, and the third simulation boosts the temperature by $10^4~$K at $z=3.5$ (which results in the hottest temperatures in void regions of the three simulations, overshooting recent estimates for the temperature; \citealt{becker11}). {\comment The simulations result in $\gamma-1$ spanning $0.41-0.58$ at $z=2.4$, consistent with the measured value of  $0.54\pm0.11$ \citep{bolton14}, where $\gamma-1$ is the power-law index of the temperature-density relation.}  We generated mock absorption spectra from each simulation.  Then, we applied our continuum fitting algorithm to the mocks, finding that the three simulations result in modest differences in our $\eta$ estimates.    These biases are discussed in the next section.

{\comment We also ran one simulation with twice the spatial resolution ($8\times$ the number of particles) to test convergence in resolution.  For this study, convergence is most important in the deep voids as it is the absorption in these regions that impact continuum placement.  We find that the simulations are not perfectly converged and that the density in the deepest voids, which are used to determine the continuum, generally overshoot by $10-20\%$ compared to the higher resolution simulation.  These differences are smaller than the differences that arise between the different thermal histories.}

\subsection{results}
\label{ss:results}

\begin{figure}
\begin{center}
{\epsfig{file=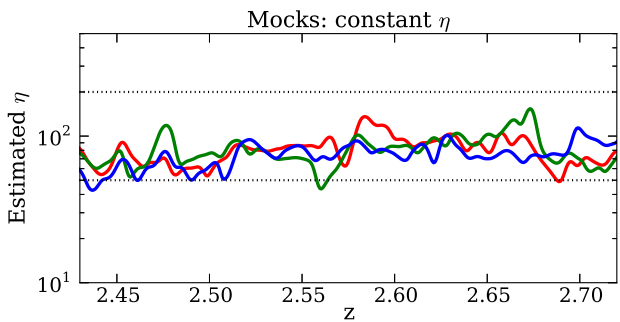, width=8.5cm}}
\end{center}
\caption{Curves show the estimated $\eta$ from three mocks that assume $\eta = 100$ and replicate the specifications of the HE2347-4342 data (aside from taking the \HeII\ Ly$\alpha$ spectra to be noiseless).  The mocks are smoothed in the same manner as in the bottom--right panel of Figure~\ref{fig:data} and are put through the same pipeline as the real data.  The factor of $\sim 2$ variation in the estimated $\eta$ owes primarily to uncertainty in the \HI\ continuum estimate.  The horizontal dotted lines show $\eta = 50$ and $200$ for reference.
\label{fig:mock_eta}}
\end{figure} 

The thick black band in the top panels in Figure~\ref{fig:data} is the normalized flux in the \HeII\ Ly$\alpha$ forest of HE2347-4342.  Its width represents the statistical error.  The spectrum in the lefthand panel shows the highest signal-to-noise span of the post \HeII\ reionization COS spectrum ($2.57<z<2.72$) binned in two pixel widths.  The spectrum in the righthand panel is binned similarly but is also convolved with a Gaussian with a FWHM of $14$ COS pixels (to reduce the noise, which is particularly large at lower redshifts) and extends over $2.43 < z < 2.72$ (which corresponds to a path length of $330$~comoving Mpc).  The lowest redshift in the specified range was chosen to exclude a geocoronal Ly$\beta$ line, and the highest to not overlap with a Gunn-Peterson absorption trough (which likely indicates a \HeII\ region and the onset of \HeII\ reionization; \citealt{shull10, mcquinn-GP}).  The horizontal error bars in the bottom panels give the size of two COS pixels (left) or the FWHM of the smoothing function (right).   

Next we multiply the UVES \HI\ Ly$\alpha$ optical depth measurement, $\tau_{\rm HI}^{\rm UVES}(z) = -\log{T_{\rm HI}^{\rm UVES}(z)}$, by the specified $\eta$ to generate the expected high-resolution \HeII\ transmission for that $\eta$ (colored curves), i.e.,
\begin{equation}
\widehat{T}_{\rm HI  \rightarrow HeII}^{\rm UVES}(z, \eta)  = \exp\left[-\frac{\eta}{4} \tau_{\rm HI}^{\rm UVES}(z) \right].
\label{eqn:THeIIeta}
\end{equation}
We use the superscript `UVES' to indicate the line-resolved spectrum at the UVES pixel size ({$\approx3.5~$km~s$^{-1}$}) and the superscript `G140L' for the line-unresolved spectrum binned to $2$ COS G140L pixels ({$\approx65~$km~s$^{-1}$}).  After $\widehat{T}_{\rm HI  \rightarrow HeII}^{\rm UVES}(\eta)$ is computed, it is then convolved with the COS line spread function (LSF; or this plus a Gaussian) and rebinned at the G140L pixel size to generate $\widehat{T}_{\rm HI  \rightarrow HeII}^{\rm G140L}(\eta)$ -- the colored thin solid curves in the top panels of Figure~\ref{fig:data}.  The COS LSF FWHM is 7 COS pixels, but with broad wings.  The statistical errors in the $S/N \approx 100$ \HI\ Ly$\alpha$ spectrum have a negligible impact on  $\widehat{T}_{\rm HI  \rightarrow HeII}^{\rm G140L}(\eta)$.  With these operations, the derived $\widehat{T}_{\rm HI  \rightarrow HeII}^{\rm G140L}(\eta)$ for the proper choice of $\eta$ should be equal to the observed \HeII\ Ly$\alpha$ transmission field, ${T}_{\rm HeII}^{\rm G140L}$, except for noise, continuum fitting errors, and the different thermal widths of \HeII\ and \HI\ lines -- systematics that we will quantify.

The comparison of the colored curves showing $\widehat{T}_{\rm HeII}^{\rm G140L}(\eta)$ with the thick black curve showing the observed \HeII\ Ly$\alpha$ transmission field, ${T}_{\rm HeII}^{\rm G140L}$, suggest that over most redshifts the observed \HeII\ spectra favor $\eta$ values between $70$ and $150$.  The bottom panels in Figure~\ref{fig:data} show ${\eta}$ estimated from tuning $\widehat{T}_{\rm HeII}^{\rm G140L}(z, \eta)$ to match ${T}_{\rm HeII}^{\rm G140L}(z)$ at each pixel.  The green regions denote the statistical uncertainty, which is calculated by matching to instead ${T}_{\rm HeII}^{\rm G140L}(z) \pm \delta {T}_{\rm HeII}^{\rm G140L}(z)$.  In the bottom--left panel, the data support a significant deviation from $\eta \approx 100$ only around $z=2.66$.  A factor of $\sim 2$ decrease in $\eta$ is also suggested at a couple locations in the $z <2.5$ data but with reduced significance (see the bottom--right panel).

\begin{figure}
\begin{center}
{\epsfig{file=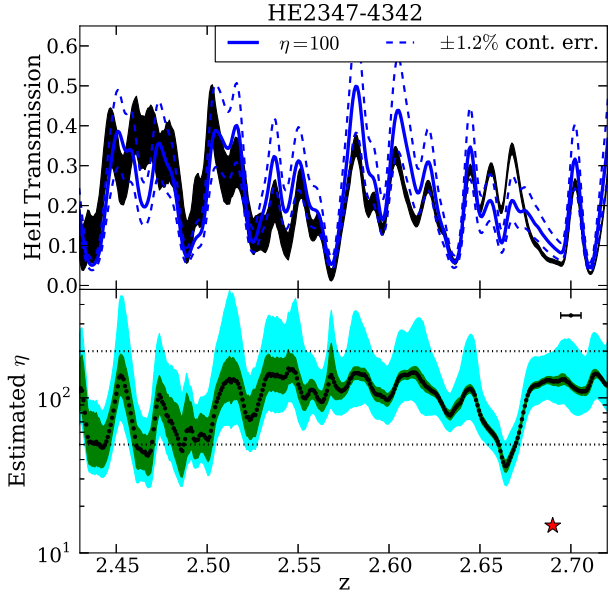, width=8.5cm}}
\end{center}
\caption{Both panels are similar to the corresponding panels on the righthand side of Figure~\ref{fig:data}, but they more explicitly account for errors in the \HI\ Ly$\alpha$ continuum estimate.  The top panel shows the derived \HeII\ transmission from the \HI\ data for a model that takes $\eta = 100$ and our estimated \HI\ continuum (blue solid curve) as well as this model except with the continuum in the \HI\ data adjusted by $\pm1.2\%$ (blue dashed curves).  This adjustment equals the RMS of the residuals of our automated continuum fitting algorithm applied to the mocks.  All but a few regions of the observed \HeII\ transmission (the thick black `curve') are consistent with $\eta = 100$ once allowing for this level of uncertainty in the \HI\ continuum. The bottom panel shows the estimated $\eta$ for our fiducial continuum fit (black dots), plus the statistical uncertainty from detector noise (green region), plus the uncertainty from $\pm1.2\%$ fractional errors in the continuum fit (cyan region).  The red star in the bottom panel shows the redshift of the only identified proximate quasar, which likely sources the $z=2.66$ feature.
\label{fig:eta_measurement}}
\end{figure} 

The curves in Figure~\ref{fig:mock_eta} represent $\eta$ estimated from three mock skewers calculated from our fiducial simulation.  These curves assume a constant $\eta = 100$ across all pixels and use the identical continuum fitting plus estimation pipeline as the real data.  The primary difference with the real data is that these mocks do not include noise in the \HeII\ Ly$\alpha$ forest (but they do match the RMS noise in the \HI\ Ly$\alpha$ spectrum).  These mocks show that our estimation procedure has intrinsic errors.  These errors owe primarily to mistakes in the continuum estimate; the different thermal line widths in the \HI\ and \HeII\ forests generally result in smaller errors (Section~\ref{sec:model}).  This figure suggests that smaller variations in $\widehat{\eta}$ than a factor of $\approx2$ can be caused by continuum misestimates.  While Figure~\ref{fig:mock_eta} used our fiducial simulation, we find a similar residual level about $\widehat{\eta}=100$ in the other two simulations, with the different simulations returning $\langle \widehat{\eta} \rangle$ that differ by $\approx 30~$per cent.\footnote{Flux calibration uncertainties are larger at the ends of the UVES echelle orders.  Flux miscalibration would lead to additional errors in the data (up to a few percent) that are not accounted for in our mocks.}

\citet{shull10} reported one order--of--magnitude variations in $\eta$ using the same COS and UVES data but processed with different reduction techniques. The grey points in the bottom panels of Figure \ref{fig:data} are the $\eta$ values estimated in \citet{shull10} with the simple estimator  
\begin{equation}
\widehat{\eta}_{\rm simple}^{\rm X}(z) = 4 \, \left[ \frac{\tau^{\rm X}_{\rm HeII}(z)}{\tau_{\rm HI}^{\rm X}(z)} \right],
\end{equation}
with X = G140L.\footnote{The smoothing used in \citet{shull10} differs somewhat from that in $\widehat{\eta}_{\rm simple}$:  \citet{shull10} convolved the UVES spectrum with the COS FWHM and then up--sampled it at the UVES resolution (which results in more data points but that are highly correlated over the COS FWHM).}
This estimator results in biased ${\eta}$ estimates because $\tau_{\rm HI}^{\rm G140L}$ is not the resolved (or true) optical depth.  The magnitude of the bias can be noted by comparing our $\eta$ estimates, the black points, to the grey \citet{shull10} estimates.  We find a similar bias (and variance) when applying $\widehat{\eta}_{\rm simple}$ to our $\eta = 100$ mocks.  To understand why $\widehat{\eta}_{\rm simple}$ is biased low when the spectrum is unresolved, note that for $\eta \geq 4$
\begin{equation}
4 \leq 4 \frac{-\log \left[\int \exp(-\eta/4 \, \tau) \, P(\tau) \,d\tau \right]}{-\log \left[\int \exp(-\tau) P(\tau) \,d\tau \right]} \leq \eta
\label{eqn:etaineq}
\end{equation}
 for any probability distribution of \HI\ optical depths in an unresolved pixel, $P(\tau)$.   The middle quantity in Eq.~(\ref{eqn:etaineq}), which is $\widehat{\eta}_{\rm simple}^{\rm X}$ in the (applicable) limit of ``pure turbulent broadening'', equals $\eta$ for $\delta$-function $P(\tau)$, and it tends to $4$ for pixels that include both large and small $\tau$.  Thus, these inequalities demonstrate why the \citet{shull10} estimates satisfy $4 \leq \widehat{\eta}_{\rm simple}^{\rm G140L} \lesssim \widehat{\eta}$.  In addition, the thin solid blue line in the bottom--left panel of Figure~\ref{fig:data} is our application of the estimator used in \citet{shull10}, $\widehat{\eta}_{\rm simple}^{\rm G140L}$, to the data put through our pipeline.  The level of agreement with the grey dots demonstrates that the differences we find with \citet{shull10} owe primarily to our improved $\eta$ estimator.

We have seen in Figure~\ref{fig:mock_eta} that the largest obstacle towards detecting fractional variations in $\eta$ of $\lesssim 50\%$ using our forward-fitting estimation method is the uncertainty in the \HI\ Ly$\alpha$ forest continuum (although the major difference with the previous results in \citealt{shull10} was our different estimator).
Figure~\ref{fig:eta_measurement} attempts to account for this additional uncertainty in the $\widehat{\eta}$ values.  The blue solid curve in the top panel shows the derived \HeII\ transmission from the \HI\ data for $\eta = 100$ and our best-fit continuum model.  The blue dashed curves assume an \HI\ continuum that differs from our best-fit estimate by $\pm1.2\%$, the RMS residual found when applying our estimation pipeline to mocks.    The bottom panel shows the estimated $\eta$ for our fiducial continuum fit (black dots), plus statistical uncertainty (green region), plus systematic uncertainty from $\pm1.2\%$ errors in the continuum fit (cyan region).  All regions aside from the feature at $z=2.66$ are roughly consistent with $\eta = 100$ when allowing for this level of uncertainty in the \HI\ continuum.  While our mocks show that factor of $2$ deviations can occur just from continuum fitting (Fig.~\ref{fig:mock_eta}), this exercise suggests that the $z=2.66$ feature does not owe to an error in the continuum.

Metal lines in the \HI\ Ly$\alpha$ could spuriously induce smaller values of $\widehat{\eta}$.   \citet{fechner07} reported the locations of metal line systems in the \HI\ Ly$\alpha$ spectrum of HE2347-4342.  We have visually inspected these regions and find little evidence for their impact at the coarse resolution of the G140L grating.  

The FUSE spectrum of HE2347-4342 has higher resolution than the COS spectrum used here, with $\lambda/\Delta \lambda = 20,000$.  This resolution makes FUSE potentially sensitive to $\eta$ on $10\times$ smaller scales than the COS G140L resolution of $2~$comoving Mpc, which could ostensibly explain why the FUSE measurements in \citet{shull04} found $10\times$ larger fluctuation amplitudes than the COS measurements in \citet{shull10}.  In addition, because of the higher resolution of FUSE, one might expect that $\widehat{\eta}_{\rm simple}$ applied to the unbinned FUSE data [as done in \citet{shull04} and \citet{fechner07}] should be less biased.  However, in practice $\widehat{\eta}_{\rm simple}$ applied to FUSE is also very biased.  The quasar continuum is usually over--fitted with the standard human fitting techniques by $\sim 2-3$ per cent at relevant redshifts (e.g., \citealt{faucher07}; we optimized our automated method here to result in smaller biases than the standard by--eye method; Fig.~\ref{fig:contfits}).  Thus, the \HI\ optical depth typically has to be larger than at least $0.04$ to not be spuriously altered by continuum fitting errors at the factor of two level.  Furthermore, once $T_{\rm HeII} < (S/N)^{-1}$ in a pixel, $\widehat{\eta}_{\rm simple}^{\rm FUSE}$ will again typically err at a factor of two.  For $\eta = 100$ (such that $\tau_{\rm HeII} = 25\, \tau_{\rm HI}$) and $S/N=5$ (characteristic of FUSE), this translates into the condition that a factor of $> 2$ error typically occurs once $\tau_{\rm HI} > 0.06$.  
  {\it Thus, only for pixels with $ 0.04< \tau_{\rm HI}<0.06$ do we expect $\widehat{\eta}_{\rm simple}$ applied to the unbinned FUSE data to result in an $\eta$ estimate that is accurate to within a factor of 2.}  We suspect that the large $\eta$ fluctuations found in \citet{shull04}, who utilized pixels with $0.02<\tau_{\rm HI} <3.9$ in the unbinned FUSE spectrum ($\Delta \lambda = 0.05$\AA), result from these issues.  \citet{shull04} also binned in $\Delta \lambda =0.2~$\AA, improving the $S/N$ per resolution element but still found similar fluctuation levels.  However,   
the maximum $\tau_{\rm HI}$ that can be used at fixed fractional error increases only as $\log(S/N)$, and, as we have seen, once the absorption becomes unresolved this leads to additional complications.\footnote{\citet{zheng04} measured $\eta$ with the FUSE data by fitting individual lines and found a comparable level of fluctuations to \citet{shull04}.  Line-fitting can be thought of as a way to regularize the optical depth field and, hence, to increase the $S/N$ on the absorption for well--resolved features.  While it likely fares better than applying $\widehat{\eta}_{\rm simple}$ to the unbinned FUSE data, it again can only be applied to pixels that span a small range of \HI\ optical depths to accurately recover $\eta$, but was applied to most pixels in \citealt{zheng04}.}

Using the FUSE data, \citet{shull04} found that $\eta$ towards HE2347-4342 was strongly correlated with voids (defined as $0.02 < \tau_{\rm HI} < 0.05$) and anti-correlated with filaments ($0.05 < \tau_{\rm HI} < 3.9$) in the coeval \HI\ Ly$\alpha$ forest.  The \citet{shull04} estimation method likely results in spurious correlations with the \HI:  Regions with low \HI\ optical depths would result in high $\eta$ estimates because of over--fitting the continuum, and those with higher \HI\ optical depths would be biased low half of the time because of the low $S/N$ of the FUSE observation (and \citealt{shull04} did not include pixels with $\tau_{\rm HeII} >2.3$, eliminating most of the filament pixels that are biased high).  These biases are consistent with the trends found in \citet{shull04}.  In our $\widehat{\eta}$ towards HE2347-4342, no obvious correlations between $\eta$ and $\tau_{\rm HI}$ are present, with the caveats that (1) at the G140L resolution, there is a small number of structures in our redshift range and (2) we are only sensitive to $\eta$ in COS pixels that include sub-regions with line-resolved optical depths of $\tau_{\rm HI} \lesssim 0.15$. 
  However, significant correlations with density are not expected in post--reionization ionizing background models (Section \ref{sec:model}).

In summary, there is little evidence for significant deviations from $\eta = 100$ towards HE2347-4342.  Uncertainty in the continuum placement results in a factor of $2$ uncertainty in $\widehat{\eta}$ in most pixels.  We hence rule out variations at the factor of $2$ level at the $2~$comoving Mpc resolution of COS (but in detail the allowed fluctuation level on these scales is smaller as the continuum errors are coherent over larger scales than the COS resolution).  The order-of-magnitude variation in $\eta$ found in previous studies owed to the use of a biased estimator (for the studies using the COS G140L data) and noisy data (for those using the FUSE data).  {\comment Our study does not rule out large fluctuations correlated over $\Delta z \gtrsim 0.1$ because of the limited path length probed by HE2347-4342.}

  \section{comparison with ultraviolet background models}
 \label{sec:model}

This section shows that the previous section's constraint on spatial fluctuations in $\eta$ is consistent with the predictions of standard ionizing background models and discusses whether the constraints rule out more exotic background models.  Section~\ref{sec:modelspecs} describes our ionizing background model, and Section~\ref{sec:Jfluconeta} investigates the impact of different ionizing background models on $\widehat{\eta}$.  We note that Appendix~\ref{sec:Lyalpha} briefly discusses how fluctuations in the \HI--ionizing background impact the small-scale \HI\ Ly$\alpha$ forest.
 
 \subsection{model for intensity fluctuations}
  \label{sec:modelspecs}
  
 \begin{figure*}
\begin{center}
{\epsfig{file=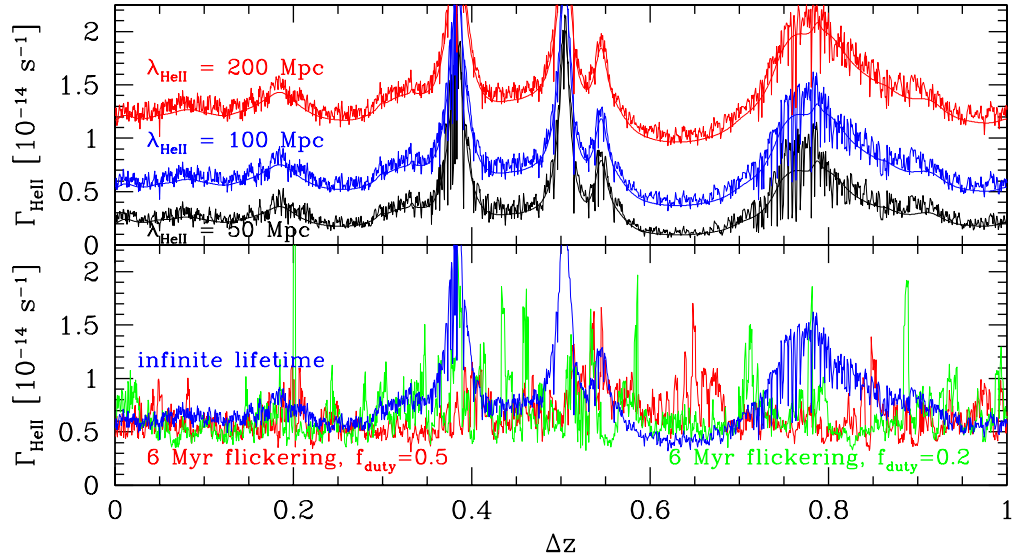, width=14cm}}
\end{center}
\caption{Skewers through our $z=2.5$ inhomogeneous \HeII\ photoionization rate models as a function of the redshift difference.  The smooth curves in the top panel (which are largely obscured by the jagged curves) are the homogeneous absorption model employed in previous studies, and the jagged curves model the radiative transfer through discrete absorption systems in the manner described in the text.  The top panel compares models in which the quasars have infinite lightbulb luminosities (with the specified $4~$Ry photon mean free paths in comoving units).  The bottom compares an infinite lightbulb model to ones in which the quasars are on for $6~$Myr, off for the next $6 \times (f_{\rm duty}^{-1} - 1)~$Myr, and the cycle repeats.  All models in the bottom panel have $\lambda_{\rm HeII} = 100~$comoving Mpc.  
  The largest spikes in both panels owe to a local enhancement from a proximate quasar.  {\it These calculations show that only a fraction of the volume deviates by a factor of $2$ from the median $\Gamma_{\rm HeII}$ even for the $\lambda_{\rm HeII}=50~$comoving Mpc case.}  The two sightlines studied in this paper both probe $\Delta z \approx 0.3$.
\label{fig:Jfluc}}
\end{figure*} 

There are three potential causes of $\eta$ fluctuations in a photoionized IGM:  the local enhancement in the backgrounds near ionizing sources, radiative transfer effects owing to absorbers of ionizing photons (i.e., the \HI\ and \HeII\ Lyman-limit systems), and the imprint of structure in the light curves of the sources.  Here we attempt to model each of these three effects in the context of models in which the sources are quasars.  Quasars are likely the dominant sources of the \HeII--ionizing background and are also a significant (if not the dominant) contributor to the \HI--ionizing background at the studied redshifts.   The discrete nature of intergalactic absorbers and the potentially complex light curves of quasars have been ignored in previous models of the ionizing background, which had modeled the fluctuations that owe to the stochastic distribution of quasars \citep{zuo92, zuo93, meiksin04, furlanettoJfluc}.  \citet{furlanetto11} speculated that the discrete nature of absorbers could be responsible for some of the features in $\eta$ found in \citet{shull04}.

Our models ignore fluctuations in the \HI--ionizing background, which are expected to be smaller by $\lambda_{\rm HeII}/\lambda_{\rm HI} \sim 0.1$, where $\lambda_X$ is the mean free path of photons at the ionization potential of ionic species $X$ (which are more relevant for the \HI\ Ly$\alpha$ forest; Appendix~\ref{sec:Lyalpha}).\footnote{\comment The fluctuations in the \HI--ionizing background are most significant in \HI\ proximity regions -- regions where the ionizing intensity from a local quasar exceeds the background value.  As the \HeII--ionizing background is also enhanced in these regions, $\eta$ in an \HI\ proximity region plateaus to $0.43\times 4^{1+\alpha}$ or $\approx 10-20$ for quasars.  However, it is much less likely (by the factor $[\lambda_{\rm HeII}/\lambda_{\rm HI}]^2$) that a skewer intersects an \HI\ proximity region, further justifying our approximation. }
  To generate a realization of the \HeII--ionizing background, we populate a cubic computational volume with randomly-placed absorbers (as described in Appendix \ref{ap:absorber_model} and briefly here).    Unlike for the \HeII\ Ly$\alpha$ absorbers, modeling the absorption systems of \HeII--ionizing photons as randomly placed discrete clouds is a decent approximation as they are much rarer:  shot noise in the number of absorbers dominates over clustering on the scale of the ionizing photon mean free path -- the scale above which attenuation is significant.\footnote{Over a skewer of length the photon mean free path -- the scale where absorption begins to matter --, the abundance of absorbers fluctuates at ${\cal O}(1)$ owing to discreteness.  Whereas, the standard deviation in their number from clustering is $<0.1 \, b$ for a mean free path of $>100~$comoving Mpc, where $b$ is the order--unity linear bias of the \HeII\ Lyman-limit systems (i.e., the sample variance fluctuations are a factor of $\gtrsim 10/b$ smaller than the Poissonian ones).}  
  Our absorber model takes one free parameter -- the mean free path of $4~$Ry photons --, which sets the abundances of the absorbers, as their cross-sectional radius is taken to be the Jeans' length.  The Jeans' length assumption is motivated in \citet{schaye01} and has tested favorably in cosmological simulations \citep{mcquinn-LL, altay11}.  In addition, we estimate the mean free path of $4~$Ry photons to be $\lambda_{\rm HeII} \approx 100 \; (\eta/100)^{-0.6}~$comoving Mpc at $z=2.5$, using standard techniques first developed in \citet[][see Appendix~\ref{ap:absorber_model}]{haardt96}.  Because our measurement in Section~\ref{ss:results} favors $\eta \sim 100$, most of our models will take $\lambda_{\rm HeII} = 100~$comoving Mpc.

Next, we place quasars randomly in a cosmological volume with luminosities drawn in a manner that reproduces the \citet{hopkins05} luminosity function.  As with the absorbers, we ignore correlations between the IGM density and the locations of quasars.  An $\cal{O}$(1) correlation with the density is only expected if a quasar falls $\sim 1~$proper Mpc from the skewer (roughly the nonlinear scale at $z=2.5$), which is unlikely, and correlations with the \HI\ transmission will be even smaller owing to the nonlinear mapping between density and $\tau_{\rm HI}$ (see \citealt{faucher08}).  The quasars either are assumed to be continuously shining lightbulbs or to flicker on and off in a manner that reproduces the luminosity function.   Such quasar variability would not be surprising \citep{kawaguchi98, collier01, hopkins10}.  Next, we trace skewers through our computational volume from all of the sources to our mock sightlines with $\sim 1~$km~s$^{-1}$ spacing, accounting for the continuum absorption of all absorbers that each skewer intersects.  We only model this direct radiation and not the re-emission of $4~$Ry photons from recombinations, which would result in a more uniform background and reduce the amplitude of the intensity fluctuations by $\sim 20\%$.

Figure \ref{fig:Jfluc} shows skewers of length $\Delta z = 1$ through several models of the \HeII\ photoionization rate at $z=2.5$.  These curves were computed using absorber models with different mean free paths (top panel) and different quasar light curve assumptions (bottom panel).
  The box size for the presented calculations is $0.9~$Gpc, and the skewers represent a randomly selected linear path through the box (with the same random numbers used between the different $\lambda_{\rm HeII}$ cases).  The smooth curves in the top panel (that fall underneath the jagged curves with the same line style) assume the attenuation is homogeneous (e.g., $d \tau = ds/\lambda_{\rm HeII}$, where $\tau$ is the optical depth and $s$ is the distance along a ray).   This attenuation model was adopted in previous studies of intergalactic intensity fluctuations.  The jagged lines show our full model, which includes absorber discreteness.  The curves in the top panel take quasars to have infinite lifetimes and lightbulb light curves, and they are computed for the specified mean free paths.  All models have the same sources and, therefore, the same ionizing emissivity.  The different overall normalization of the $\Gamma_{\rm HeII}$ curves owes to the variation in the mean free path.

There are a few properties of $\Gamma_{\rm HeII}$ (and $\eta$) that one should note from the top panel of Figure~\ref{fig:Jfluc}.  First, the largest fluctuations are from quasar proximity regions -- regions in which the flux is enhanced by a factor of $2$ over the background.  Proximity regions fill $ (6 \sqrt{\pi})^{-1} n^{-1/2} \,\lambda_{\rm HeII}^{-3/2}$ of the volume, where $n$ is the luminosity squared--weighted source number density.  Quasars are the rarest objects in the Universe that can dominate the ionizing background.  Thus, for other source models (which have larger $n$), the proximity regions constitute an even smaller fraction of space and the $\Gamma_{\rm HeII}$ fluctuations are correspondingly smaller.  Second, the fraction of space (in a quasar proximity region) that occupies a factor of $2$ excursion above the background (or median) $\Gamma_{\rm HeII}$ is small for all the considered quasar models, scaling as $\propto \lambda_{\rm HeII}^{-3/2}$.  Even for the curve that represents our shortest mean free path model with $\lambda_{\rm HeII} = 50~$comoving Mpc, a small fraction of the skewer falls a factor of $2$ above the mean.  Thus, even such short mean free path models could not reproduce the old $\eta$ measurements that found $1~$dex fluctuations, contrary to the claim in \citet{bolton06}.  Third, the inhomogeneous distribution of absorbers rarely leads to fluctuations that are larger than $10$s of per cent (compare the jagged and smooth curves in the top panel).

{\comment
While our models do not account for dispersion in $\alpha$, the ultraviolet spectral index of $f_\nu$ [erg~$cm^{-2}$~s$^{-1}$~Hz$^{-1}$~sr$^{-1}$], such dispersion is unlikely to change the character of fluctuations in $\eta$ \citep{furlanetto09}.  The RMS amplitude of fluctuations is $\propto \sqrt{\langle L^2 \rangle_Q/ [ \bar{n}_{\rm QSO} \langle L \rangle_Q^2]}$, where $\langle ... \rangle_Q$ denotes an average over the quasar luminosity function and $\bar{n}_{\rm QSO}$ is the luminosity--weighted quasar number density \citep{zuo92, meiksin04}.  Therefore, only if the dispersion in $\alpha$ correlates with the $4~$Ry luminosity ($L$) -- altering the $4~$Ry luminosity function relative to our input model at $1$Ry -- is the RMS fluctuation level affected.  

If the faint-end of the luminosity function is softer or bright-end slope is harder than in our fiducial model, the fluctuation amplitude increases by creating larger quasar proximity regions.  However, these changes to the luminosity function result in more large-scale ($\delta z > 0.05$) variations rather than creating the $\sim 1~$dex intensity variations on Mpc--scales suggested by previous $\eta$ observations.  The fluctuation amplitude is also somewhat sensitive to the maximum and minimum luminosities at which one evaluates the luminosity, which we took to be bolometric luminosities of $10^{42}$ and $10^{48}~$erg~s$^{-1}$.   Factors of $10$ changes in these cutoff luminosities result in factor of $\sim 1.5$ changes in the RMS intensity, but again with this primarily affecting fluctuations  correlated over $\delta z > 0.05$.  
}

 \begin{figure*}
\begin{center}
{\epsfig{file=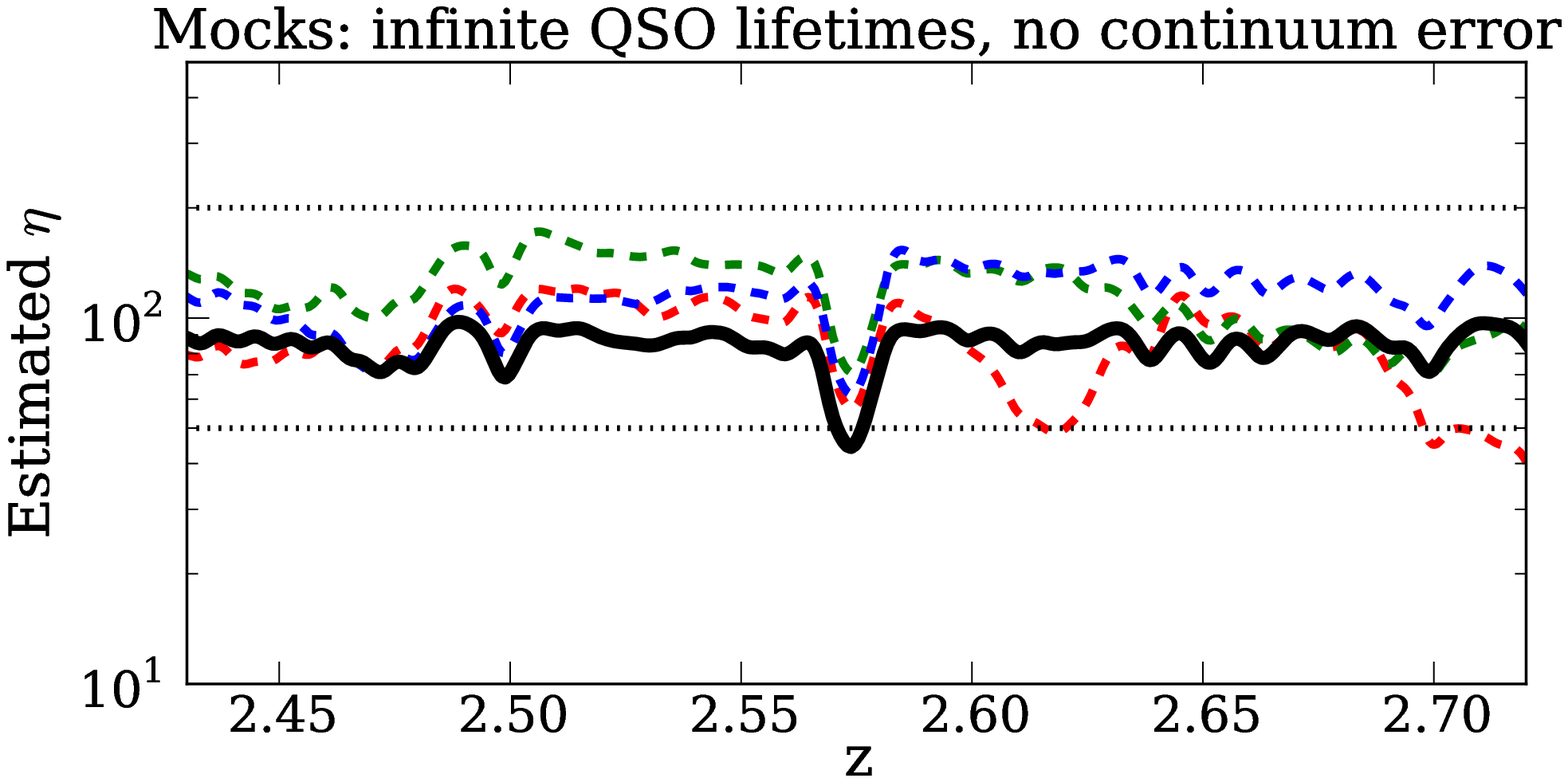, width=8cm}}
{\epsfig{file=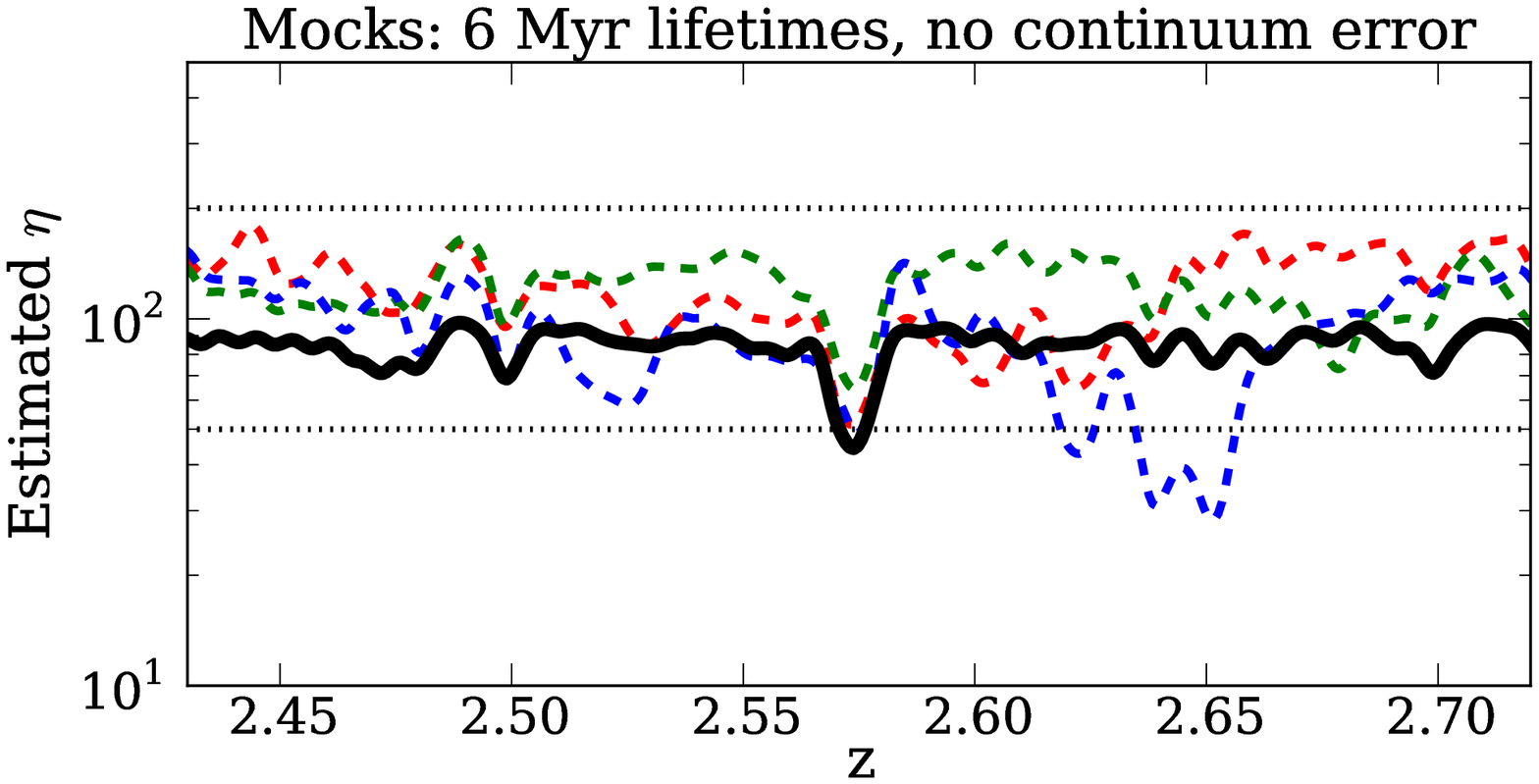, width=8cm}}
{\epsfig{file=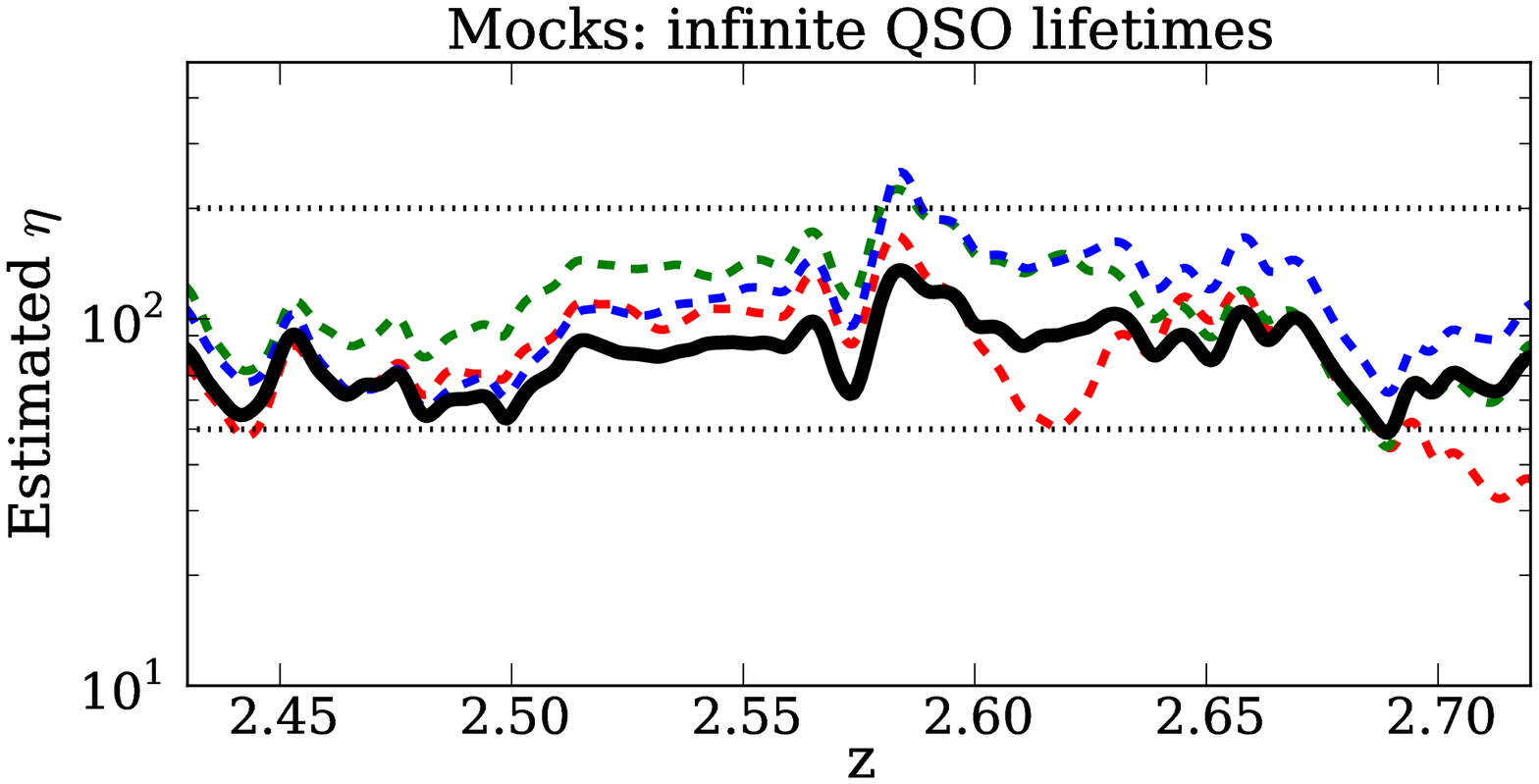, width=8cm}}
{\epsfig{file=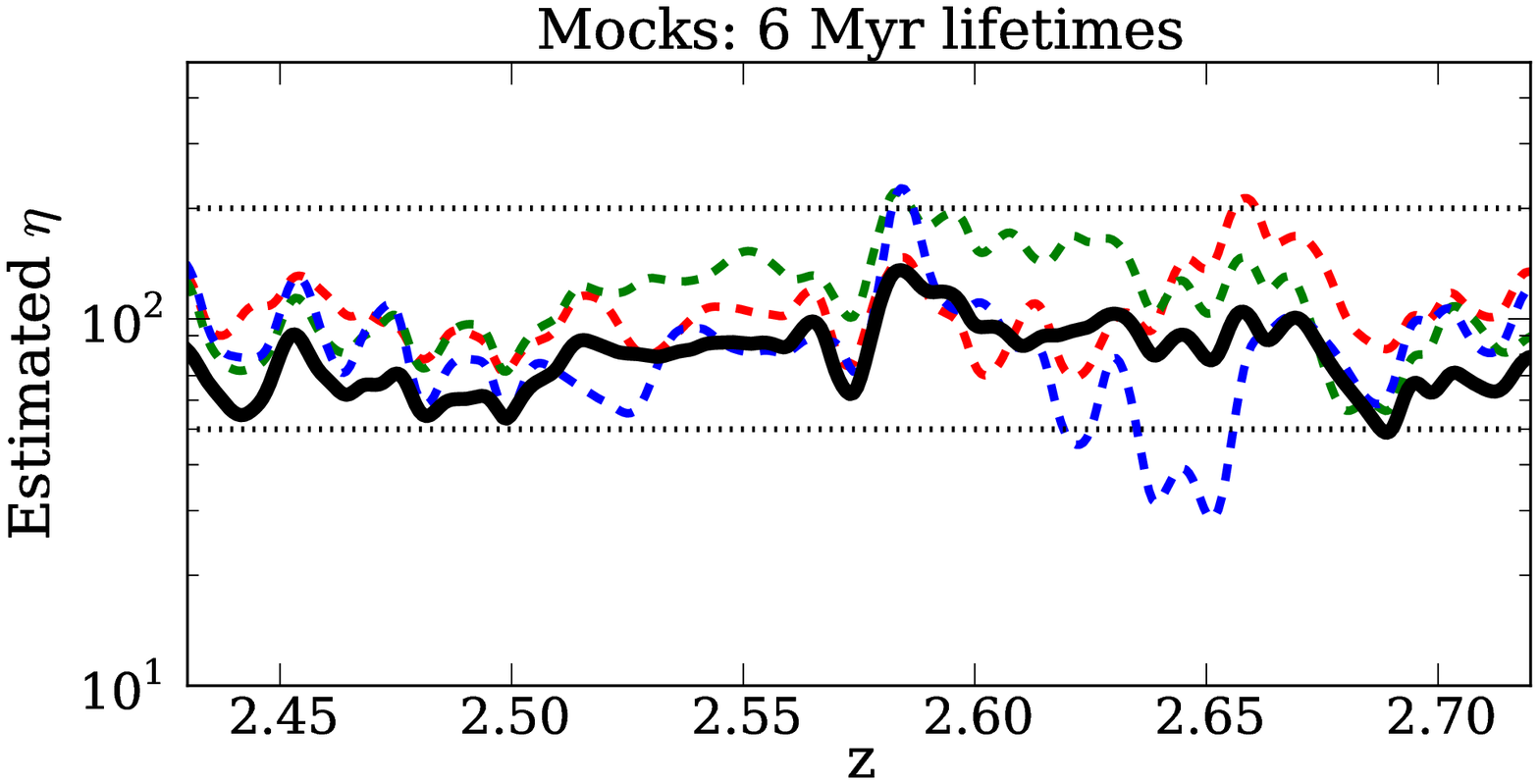, width=8cm}}
\end{center}
\caption{Impact of spatial fluctuations in $\Gamma_{\rm HeII}$ on $\widehat{\eta}$ for noiseless mocks with $\langle \eta \rangle = 100$, $\lambda_{\rm HeII} = 100~$comoving~Mpc, and specifications matching the HE2347-4342 data.  The black solid curve in each panel shows the case without $4~$Ry background fluctuations, and each of the colored dashed curves represents a different realization of the $4~$Ry intensity field.  The colored dashed curves in the left panels assume quasars with infinite, lightbulb light curves; those in the right panels assume light curves in which the quasars flicker on and off with a duty cycle of $6~$Myr, which we argue in the text results in maximal fluctuations.  The top panels show the idealized case in which the continuum is known -- the fluctuations in the black solid curves owe to the different thermal widths of hydrogen and helium --, and the bottom panels the case in which the continuum is estimated with the same algorithm (and $S/N$ ratios in the \HI\ spectrum) as the HE2347-4342 data.  The curves in the bottom panels should be compared with the measurement presented in the bottom panel of Figure~\ref{fig:eta_measurement}; the curves have been smoothed to the identical resolution as in Figure~\ref{fig:eta_measurement}.  The level of statistical fluctuations seen in either model is qualitatively consistent with the allowed level of fluctuations in $\widehat{\eta}$ towards HE2347-4342.
\label{fig:etaJmocks}}
\end{figure*}    
  
{\comment
  Only fluctuations in quasar light curves (which do not alter the RMS fluctuation level at fixed luminosity function) \emph{can} imprint sizable Mpc--scale $\eta$ variations.}  
  The bottom panel shows again the $\lambda_{\rm HeII}=100~$comoving Mpc, infinite--lifetime, lightbulb case as well as two cases in which the quasars flicker on a timescale of $6~$Myr:  One in which each quasar is on half of the time ($f_{\rm duty} = 0.5$; red curve) and another with $f_{\rm duty} = 0.2$ (green curve).  The number of quasars is adjusted in the flickering cases to reproduce the observed quasar luminosity function.  Both flickering models result in ``maximal'' Mpc--scale $\Gamma_{\rm HeII}$ fluctuations for the reasons detailed in the ensuing footnote.\footnote{Flickering on a few~Myr timescales results in fluctuations that are comparable to the scale resolved by the COS G140L grating, and $6~$Myr is similar to the photoionization equilibrium timescale, $\Gamma_{\rm HeII}^{-1}$ -- the timescale over which the \HeII\ ionization state adjusts to a new background level.  
    Thus, $6~$Myr flickering quasar models roughly maximize the level of $\Gamma_{\rm HeII}$ fluctuations that are possible.  We also note that even if the \HeII\ were out of equilibrium, it would appear to have an intermediate value of $\eta$ between its last and next equilibrium state, and this intermediate value would not depend on density.}  
 
It is difficult to illustrate the character of $\eta$ fluctuations in a more quantitative plot than Figure~\ref{fig:Jfluc}.   Probability distribution functions of $\Gamma_{\rm HeII}$ hide all the small-scale fluctuations from discrete absorption systems and complex quasar light-curves -- the novel aspects of our model -- as the fluctuations that owe to source discreteness dominate the RMS.  The probability distribution functions of our $\Gamma_{\rm HeII}$ models are similar to those calculated assuming homogeneous attenuation in \citet[their Fig.~1]{furlanetto09}.

\subsection{the impact of intensity fluctuations on $\widehat{\eta}$}
\label{sec:Jfluconeta}

 Figure \ref{fig:etaJmocks} shows the impact of two intensity fluctuation models with $\lambda_{\rm HeII} = 100~$comoving Mpc on $\widehat{\eta}$ using our estimation method, where the mean $\Gamma_{\rm HeII}$ in both models has been renormalized so that $\langle \eta \rangle = 100$.  The black solid curves mark the case without intensity fluctuations, and each of the colored dashed curves represents a different realization of the intensity field.  The colored dashed curves in the left panels take infinite, lightbulb quasar light curves, and the corresponding curves in the right panels take quasar light curves in our maximal fluctuation model where the quasars flicker on and off every $6~$Myr ($f_{\rm duty}= 0.5$; the red curve in Fig.~\ref{fig:Jfluc}).  We again have not included noise in the \HeII\ Ly$\alpha$ mock skewers, and these estimates have been smoothed to the same resolution as the HE2347-4342 data in Figure~\ref{fig:eta_measurement}.  We find the range of fluctuations in the three examples given in each panel to be qualitatively representative of a larger ensemble of skewers.
 
The top panels in Figure~\ref{fig:etaJmocks} show the case in which the true continuum is known.  The fluctuations in the black solid (homogeneous background) curves owe to the different thermal widths of hydrogen and helium.\footnote{The central dip in the black solid curves in Figure~\ref{fig:etaJmocks} is quite anomalous in our mocks.  Nevertheless, the dip is still not nearly as broad as the feature at $z=2.66$ we find towards HE2347.}  The bottom panels show the case in which the continuum is estimated in the same manner as the data.  Note that all the curves on Figure~\ref{fig:etaJmocks} use different realizations of $\eta$ but are calculated from only one realization of the \HI\ Ly$\alpha$ forest. (Fig.~\ref{fig:mock_eta} shows how the residuals vary between different realizations of the \HI\ for the constant $\eta$ case.)  The bottom panels in Figure~\ref{fig:etaJmocks} should be compared with our measurement in the bottom panel of Figure~\ref{fig:eta_measurement}.  Such a comparison would favor the model that is a better match to the number and amplitude of $\eta$ fluctuations in the data.  However, the models in the lefthand and righthand panels yield a similar amplitude for the largest deviations, with the number of extrema being somewhat larger in the righthand panels.  However, it is evident that the data do not have the statistical power to confidently rule out either model for intensity fluctuations.  Because of the limited path length probed by our $\eta$ estimates towards HE2347-4342, a more quantitative statistical analysis does not seem warranted.
 
 \citet{bolton06} claimed that the level of $\widehat{\eta}$ fluctuations towards HE2347-4342 reported in \citet{zheng04} is too large to be sourced by models in which most of the \HeII--ionizing photons originate from $\sim10^6~$K shocked gas within galactic halos \citep{miniati04}.  Instead, \citet{bolton06} argued that such sources would lead to smaller fluctuation levels.  Because our measurement is
  consistent with a nearly homogeneous field, we disagree that the data support this conclusion.  Similarly, our measurement cannot rule out significant contributions to the $4~$Ry background from stars and hot gas within galaxies \citep{furlanetto08}.  We are not sensitive to the $\eta \sim 10$ that some exotic sources could produce in their proximity regions \citep{venkatesan03}, as the proximity region widths of galactic sources should be narrower than the $2~$comoving Mpc resolution of the G140L grating. 
  
Lastly, we note that the small-scale intensity fluctuations that owe to absorber discreteness have little impact on $\eta$ at the resolution of COS; the fluctuations that are most apparent in Figure~\ref{fig:etaJmocks} owe to quasar proximity regions.

\section{other implications of our $\eta$ measurement}
\label{sec:implications}

Our measurements of $\eta$ in Figure~\ref{fig:eta_measurement} are consistent with $\eta \approx 100$, aside from at $z\approx 2.66$ (see Fig.~\ref{fig:eta_measurement}).  $\langle \eta \rangle \approx 100$ is comparable to previous measurements at $z=2.5$ \citep{shull04, fechner07,worseck11}, but larger than the value found in other studies \citep{zheng04, bolton06}.\footnote{Note that the mean $\eta$, $\langle \eta \rangle$, is $\sim 30\%$ higher than the median in quasar source models (see Fig.~\ref{fig:etaJmocks} and \citealt{furlanetto09}).}  However, in many of these previous studies it was unclear how to measure $\langle \eta \rangle$ in the presence of large fluctuations, and in fact many of these studies reported the logarithmically averaged $\eta$.
 Using $\Gamma_{\rm HI} = (0.5-1) \times 10^{-12}~$s$^{-1}$ (which spans the range of recent estimates using the flux decrement method over these redshifts; e.g., \citealt{bolton05, faucher07}), $\langle \widehat{\eta} \rangle \approx 100$, and the formula $\langle \Gamma_{\rm HeII} \rangle \approx 0.43 \times \langle \Gamma_{\rm HI} \rangle / \langle \eta \rangle$ results in an estimate for the mean ${\Gamma}_{\rm HeII}$:
\begin{equation}
\langle \widehat{\Gamma}_{\rm HeII} \rangle (z) = (2-4)\times 10^{-15}~{\rm s}^{-1} ~~~~{\rm for ~~} 2.4 <z < 2.7. 
\label{eqn:GHeII}
\end{equation}
Eq.~(\ref{eqn:GHeII}) implies a long photoionization equilibrium time of $t_{\rm eq} = \Gamma_{\rm HeII}^{-1} = 8 - 16~$Myr.  Note that when exposed to a new background of  $\Gamma_{\rm HeII}$, the \HeII\ fraction adjusts as
\begin{equation}
x_{\rm HeII}(t) = x_{\rm HeII, eq} + \left[x_{\rm HeII}(0) - x_{\rm HeII, eq}\right] \, \exp \left(-t/ t_{\rm eq} \right), \nonumber
\end{equation}
where the equilibrium fraction scales as $x_{\rm HeII, eq} \propto \Gamma_{\rm HeII}^{-1}$.  Thus, detecting any quasar proximity region in the \HeII\ Ly$\alpha$ forest results in an interesting constraint on quasar lifetimes.  

Our $\widehat{\eta}$ values towards HE2347-4342 are suggestive of a \HeII\ transverse proximity zone at $z=2.66$ extending over $\Delta z \approx 0.03$ or $\approx 30~$comoving Mpc.  One proximity region over $2.3 < z < 2.7$ is also consistent with the expected number in our mocks.  If this proximity region were associated with a known quasar, it would give a direct constraint on the lifetime or beaming angle of quasars.  \citet{worseck07} identified a quasar (B-band magnitude of $20.2\pm0.2$) with an estimated redshift of $z=2.690$ using OI and CII and a flux of $29~\mu$Jy at the hydrogen Lyman-limit.  This quasar has a transverse proper separation of $r_\perp = 30~$comoving Mpc from the HE2347-4342 sightline.  \citet{worseck07} identified no other proximate quasars over the redshift range considered here when searching within $\approx 40~$comoving Mpc.  We do not see an decrease in $\eta$ at the redshift of this quasar, but only the $z=2.66$ feature centered $30~$comoving Mpc in front of it (at a $45~$degree angle).  If sourced by this quasar, such an angle suggests that the quasar has been on for $(r - r_\parallel)/c \approx 10~$Myr, where $r_\parallel$ is the line-of-sight distance from the quasar to the start of the proximity region ($\approx 20~$comoving Mpc).    Alternatively, this redshift offset could indicate that the emissions are beamed \citep{furlanetto11}.\footnote{\comment The line of sight velocity offset between the estimated quasar redshift and the ostensible transverse proximity zone is $\approx 2000\,$km/s, which is several times larger than the typically redshift error using optical lines \citep{shen11}.  Thus, quasar redshift errors are unlikely to impact our qualitative conclusions.}

  The shape of the putative proximity region supports the hypothesis that it owes to this quasar -- peaking towards the higher redshifts and falling off with distance from the quasar.  (These trends are most apparent in Fig.~1.)  Both the finite lifetime and beaming angle interpretation require a quasar lifetime of $\geq 10~$Myr owing to light travel delays.  There is additional evidence to support the finite lifetime hypothesis:  The high-redshift tail of this feature -- with a width $dz \approx 0.01$ corresponding to a light travel time difference from the quasar of $\approx 3~$Myr -- is comparable to the width expected as the \HeII\ comes into photoionization equilibrium over $4-8~$Myr (assuming a $2\times$ enhancement in $\Gamma_{\rm HeII}$).

The ${\cal O}$(1) enhancement in the \HeII\ photoionization rate as indicated by our $\widehat{\eta}$ is also consistent (given uncertainties) with being sourced by the $z=2.690$ quasar.  Assuming a single power-law for the quasar spectrum, a contribution from the quasar of
\begin{equation}
\Gamma_{\rm HeII}^{\rm quasar} = \frac{\sigma_{\rm HeII} \, f_{\nu}(E_{\rm obs})}{\left(3 + \alpha \right) \left(1+z \right) h_p} \left( \frac{E_{\rm HeII}}{E_{\rm obs} (1+z)} \right)^{-\alpha}  \frac{d_L(z)^2}{r^2}
\label{eqn:proximity}
\end{equation}
is expected, 
where $d_L$ is the luminosity distance, $f_{\nu}$ is the observed specific flux at energy $E_{\rm obs}$, $h_p$ is the Planck's constant, and $\sigma_{\rm HeII}$ is the \HeII\ photoionization cross section at $E_{\rm HeII} = 4~$Ry (and we have assumed that it scales above this frequency as $\nu^{-3}$).  
  For the $z=2.690$ proximate quasar, Eq.~(\ref{eqn:proximity}) yields $\Gamma_{\rm HeII}^{\rm quasar} = 1.2 \times 10^{-14}~$s$^{-1}$ at $z=2.66$ using the power-law of $\alpha = 0.24$ inferred in \citet{worseck07} around $1~$Ry to extrapolate from the Lyman-limit of hydrogen.  This value overshoots the enhancement that is seen, which would be mitigated if the spectrum softened at higher frequencies or if the quasar was on-average dimmer than at present over a time $\sim \Gamma_{\rm HeII}^{-1}$.  The enhancement becomes $1.5\times10^{-15}~$s$^{-1}$ if we instead use the average spectral index of quasars of $\alpha = 1.6$ \citep{telfer02}, which undershoots somewhat.  The anticipated $\Gamma_{\rm HeII}^{\rm quasar}$ would be $2.5\times$ larger at the coeval redshift of $z=2.690$, where no decrease in $\eta$ is noted.   Since quasar lightcurves are not lightbulbs, these estimates simply motivate that the enhancement that is seen is consistent with being sourced by the said quasar.\footnote{While we have shown that the $z=2.690$ proximate quasar can source this proximity region, there are two other possibilities:  (1) A quasar at $z=2.66$ below the detection limit of the \citet{worseck07} survey, corresponding to a factor of $\approx5$ lower fluxes than the $z=2.690$ proximate quasar.  Somewhat deeper searches would rule out this hypothesis as the proximity region size scales as luminosity to the $1/2$ power;  (2) a beamed quasar where the open angle does not include our line-of-sight.}

We also find tentative evidence for proximity regions towards HS1700+6416 associated with the two proximate quasars (Appendix~A).  These can be added to the detection of a proximity region associated with a proximate quasar towards QSO0302-003 at $z=3.05$ \citep{heap00, jakobsen03}.  All four of these associations suggest quasar lifetimes of $\gtrsim 10^7$~yr.  These are the most robust direct constraints on Salpeter--timescale quasar lifetimes.  (See \citealt{martin04} for a census of methods to constrain quasar lifetimes.)\\  

 The mean value of $\eta$ also translates into a constraint on the effective spectral index between the \HI\ and \HeII\ ionizing backgrounds (e.g., \citealt{miralda90, shull04, bolton06, faucher08}).  We use the parameterization $\epsilon_\nu \propto \nu^{-\alpha_{\rm eff}}$, where $\epsilon_\nu$ is the average emissivity per unit frequency.  Our measurement of $\eta$ as well as recent work to better estimate the mean free path of \HI\ and \HeII--ionizing photons \citep{prochaska09, davies12, omeara13} allows us to improve upon previous estimates for $\alpha_{\rm eff}$.  In particular, we find for the spectral index of the ionizing sources
\begin{equation}
\alpha_{\rm eff} = 1.92 + \log_4 \left [\left(\frac{0.5}{\delta} \right) \left(\frac{8}{\lambda_{\rm HI}/\lambda_{\rm HeII}} \right)\left(\frac{\eta}{100}\right)  \right],
\label{eqn:alpha}
\end{equation}
where we have taken the ratio of the hydrogen and helium recombination rates to be $5.5$ and ignored the factor $(3 + \alpha_{1 \rm Ry})/(3 + \alpha_{4  \rm Ry})$ in the log;  $\alpha_{X}$ is the true (IGM filtered) spectral index in specific intensity around energy $X$ ($J_\nu \propto \nu^{-\alpha_X}$ for $\nu \sim X$).  
  Eq.~(\ref{eqn:alpha}) is derived from Eq.~(\ref{eqn:etarels}) and using that for $\delta =1$
\begin{equation}
\langle \Gamma_{X} \rangle =  \int d\nu \, \sigma_X(\nu) \epsilon(\nu) \lambda(\nu),
\label{eqn:GX}
\end{equation}
where $\epsilon(\nu)$ is the sources' specific photon emissivity, $\sigma_X$ is the photoionization cross section, and $\lambda(\nu) \equiv ds/d\tau_{\rm eff}(\nu)$ is the frequency dependent mean free path and $\tau_{\rm eff}$ the effective photoelectric optical depth \citep{haardt96}.  

The chosen ratio of the \HI\ to \HeII\  Lyman-limit photon mean free path, $\lambda_{\rm HI}/\lambda_{\rm HeII} = 8$, in Eq.~(\ref{eqn:alpha}) is consistent with the ratio one estimates from combining recent estimates based on the \HI\ column density distribution with our calculations: $\lambda_{\rm HI}$ was recently measured to be $830\pm150~$comoving Mpc at $z=2.4$ \citep{omeara13} using the novel stacking technique developed in \citet{prochaska09}, and we estimate $\lambda_{\rm HeII} = 80-110~$comoving Mpc for $\eta = 100$ (see Fig.~\ref{fig:mfp} and Appendix~\ref{ap:absorber_model}).\footnote{
In the model described in Appendix~\ref{ap:absorber_model}, which physically relates these mean free paths to the profile of absorption systems, the ratio of the \HI\ to \HeII\  Lyman-limit photon mean free path is 
\begin{equation}
\frac{\lambda_{\rm HI}}{\lambda_{\rm HeII}} = \left( \frac{\eta}{4} \right)^{\beta - 1} \approx  8 \left( \frac{\eta}{100} \right)^{0.65},
\label{eqn:LoverL}
\end{equation}
where $\beta$ is the power-law index of the \HI\ column density distribution for systems that are optically thin at $1~$Ry;  $\beta$ has been recently constrained over relevant columns of $N_{\rm HI} ~ 10^{15}-10^{17}$ cm$^{-2}$ to be $\approx 1.65\pm0.03$ by \citet[for their fit to columns of $N_{\rm HI} >10^{14}~$cm$^{-2}$]{rudie13}, which we used to write the rightmost expression in Eq.~(\ref{eqn:LoverL}).  Plugging Eq.~(\ref{eqn:LoverL}) into Eq.~(\ref{eqn:alpha}) would yield a weaker scaling of $\alpha_{\rm eff}$ with $\eta$.}
Larger $\lambda_{\rm HI}/\lambda_{\rm HeII}$ would require a harder $\alpha_{\rm eff}$.

{\comment In detail, Eq.~(\ref{eqn:GX}) should account for redshift evolution as the time for a photon to travel $\lambda_{\rm HI}$ is a significant fraction of the Hubble time.  In Eq.~(\ref{eqn:alpha}), $\delta$ is a factor that corrects for this effect.  We find $\delta = 0.5-0.6$ if we solve the full redshift independent equations (c.f. \citealt{haardt96}) for the emissivity history of quasars assuming the redshift dependence of $\lambda_{\rm HI}$ from \citet{fumagalli13}.\footnote{We find that $\Delta \delta\sim 0.1$ results from evolution of the quasar emissivity, and the rest to the \HI\ mean free path decreasing with increasing $z$.}}

Thus, taking our favored values of $\eta=100$, $\lambda_{\rm HI}/\lambda_{\rm HeII}=8$ and $\delta = 0.5$, current data prefer an $\alpha_{\rm eff}=1.92$, which is somewhat softer with the spectral index of quasars as measured in \citet{telfer02}, who find $\alpha_q = 1.76 \pm 0.12$ for the spectral index of the composite spectrum of their full sample (in coincidental agreement with our $\alpha_{\rm eff}$ for our fiducial parameter choices), and $1.57\pm0.17$ when just including radio-quiet quasars (which constitute $\sim 80\%$ of quasars).  Recently, \citet{shull12} reported a somewhat harder (but consistent) value of $\alpha_q = 1.41\pm0.21$.  If quasars contribute $f_q$ of the \HI--ionizing background and all of the \HeII--ionizing background, their spectral index must be $\alpha_q = \alpha_{\rm eff} +  \log_4 f_q$.  For example, if half of the \HI\ background owes to stars (which should only contribute photons to the \HI--ionizing and not the \HeII--ionizing background) and half to quasars, then $\alpha_q = \alpha _{\rm eff}- 1/2$ or $\alpha_q =1.42$ for our fiducial value of $\alpha_{\rm eff}=1.92$, which is somewhat harder than the \citet{telfer02} indices but consistent with the index reported by \citet{shull12}. 
 In conclusion, it would be difficult to accommodate much more than a $50\%$ contribution to the $1~$Ry background from starlight, and the effective spectral index of the sources that we derive is $\sim 2\sigma$ consistent with the background owing to just radio quiet quasars. 

\section{Conclusions}

We reanalyzed HST data from the brightest and most studied \HeII\ sightline, HE2347-4342, finding no evidence at $2.4 < z < 2.7$ for large fluctuations in the ratio of the $1~$Ry background to the $4~$Ry background, $\eta$.  We also measured $\eta$ towards the second brightest sightline, HS1700+6416, and reached a similar conclusion (Appendix~A).   We constrain the RMS fluctuation amplitude to be $<2$ when smoothed over a proper Mpc.  In addition, we do not find any evidence for the extremely low $\eta$ values ($\sim 10$) noted in the previous studies and that may be indicative of harder source populations such as Population III stars (e.g., \citealt{venkatesan03}).

Previous studies had reached the opposite conclusion that the fluctuations were quite large.  This difference owes to three improvements in our analysis:  (1) properly treating $-\log$ of the transmission from medium resolution \HeII\ spectra as an effective optical depth rather than the physical optical depth;  (2) placing the continuum in a manner motivated by cosmological simulations; and (3) quantifying the systematic uncertainty by applying the same analysis pipeline to realistic mocks.  For the analysis of the COS G140L data discussed here, improvement (1) was the main driver of the contrasting conclusion with \citet{shull10}.  We argued that the large $\eta$ fluctuations previously found using the unbinned FUSE \HeII\ Ly$\alpha$ forest data owed to continuum errors as well as the low pixel signal-to-noise of these observations. 

We generated mock spectra with ultraviolet background models motivated by quasars being the $4~$Ry background sources.  We found that these models were consistent with our $\eta$ estimates.  In addition, we investigated models with flickering quasars (tuned to maximize $\eta$ fluctuations) and discussed the character of $\eta$ fluctuations in models in which more exotic sources (such as massive stars or hot intra-halo gas) are responsible for a significant fraction of the $4~$Ry background.  All models predict relatively small $\eta$ fluctuations that are consistent with our analysis.
 
Two other implications of our $\eta$ measurement were discussed.  First, we found evidence for one \HeII\ transverse proximity region over the surveyed redshift range in the spectrum of HE2347-4342.  We showed that the only proximate quasar within this redshift interval, which is offset in redshift by $\Delta z= 0.03$ from this feature, appears to be sufficiently luminous to be responsible.  The redshift offset would then imply that the quasar turned on $10~$Myr ago.  Another interpretation is that the offset owes to the quasar's emissions being beamed (which would still require an age of $\geq 10~$Myr owing to light travel delays).  
  This quasar--associated feature, plus our more tentative associations towards HS1700+6416 and the one reported in \citet{jakobsen03}, brings the tally to four of putative \HeII\ proximity regions that suggest $\gtrsim 10~$Myr quasar lifetimes.  
Second, we showed that our estimates for the mean $\eta$ (combined with recent improvements in \HI\ column density distribution measurements) are inconsistent with models in which stars contribute much more than half of the $z=2.5$ \HI--ionizing background and consistent at $2\sigma$ with quasars being the only source of this background.\\

We thank Cora Fechner for kindly providing her reduction of the Keck HIRES spectrum of HS1700.  We also thank Joseph F. Hennawi, Steven R. Furlanetto, Adam Lidz, J. Xavier Prochaska, and the anonymous referee for helpful comments that improved the manuscript.  MM acknowledges support by the National Aeronautics and Space Administration
through Hubble Postdoctoral Fellowship awarded by the Space
Telescope Science Institute, which is operated by the
Association of Universities for Research in Astronomy,
Inc., for NASA, under contract NAS 5-26555.
GW is supported by NASA.

\bibliographystyle{apj}
\bibliography{References}

\appendix

 \section{A: HS1700+6416}
 \label{sec:HS1700}
 
 \begin{figure}
\begin{center}
{\epsfig{file=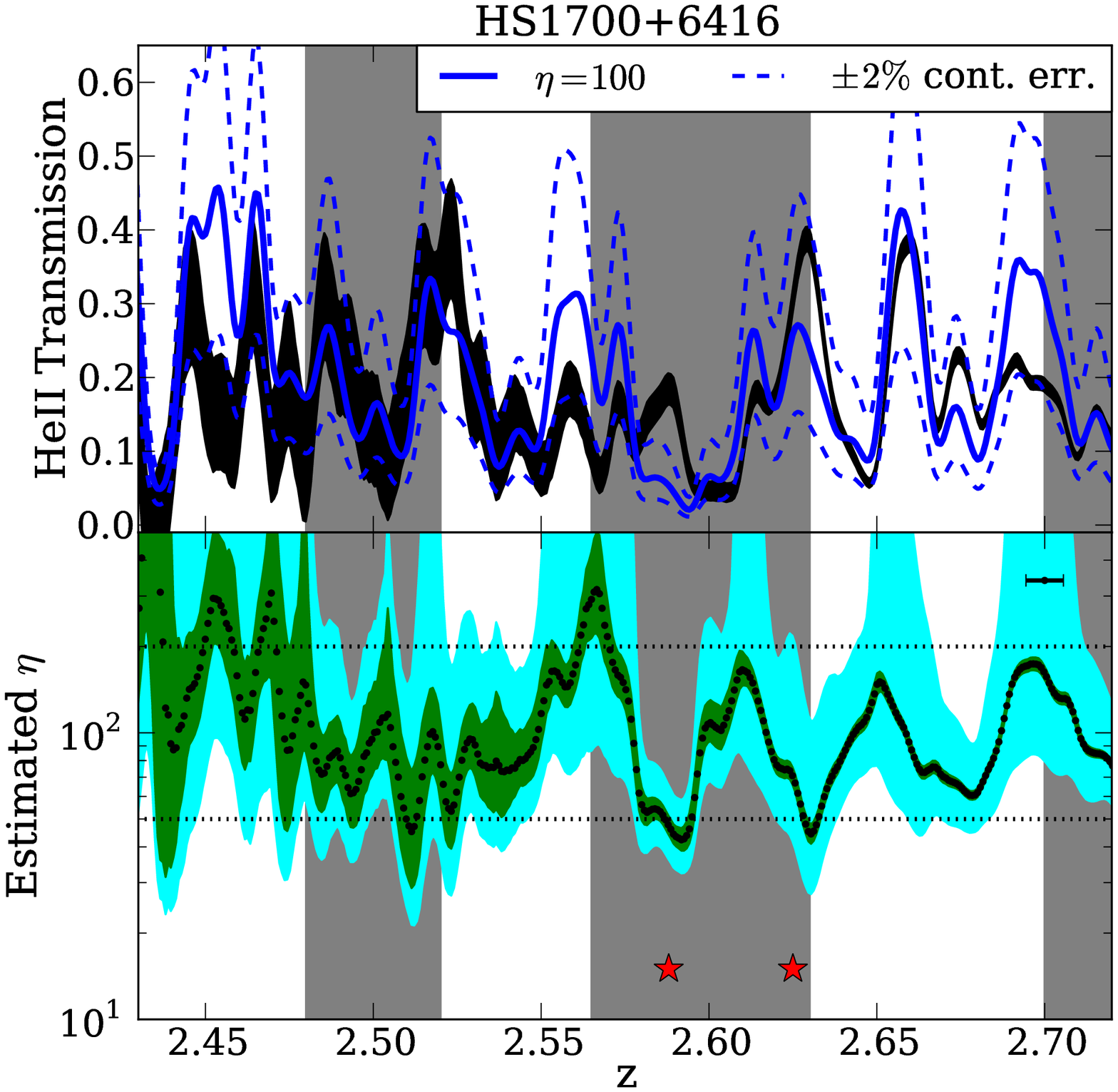, width=8.5cm}}
\end{center}
\caption{Similar to Fig.~\ref{fig:eta_measurementHS1700} except for the quasar HS1700+6416 and for the following additional differences:  The blue dashed curves in the top panel and the cyan regions in the bottom account for a continuum error of $2\%$, in line with the error in our HS1700+6416 mocks.  The grey highlighted redshifts demarcate putative continuum features in the \HI\ spectrum that may result in larger errors in the estimated continuum.  The red stars in the bottom panel show the locations of the proximate quasars identified in \citet{syphers13}.
\label{fig:eta_measurementHS1700}}
\end{figure} 
 
 The second brightest \HeII\ Ly$\alpha$ quasar is HS1700+6416 ($z= 2.75$).  The \HeII\ absorption towards this quasar has been studied previously in several papers \citep{fechner06, fechner07, syphers13}.  We again use the \HeII\ Ly$\alpha$ data taken with the COS~G140L grating.  The \HI\ Ly$\alpha$ data have $S/N\sim 80$ and were taken with Keck HIRES spectrograph, $\lambda/\Delta \lambda \approx 40,000$ \citep{fechner06}.   The continuum of HS1700+6416 has more apparent structure in it than the extremely smooth continuum of HE2347-4342.  (Metal line absorption is also more severe in HS1700+6416, with $7$ intervening Lyman-limit systems; e.g., \citealt{fechner06}.)  As a result of the continuum structure, it was less clear where the continuum should be placed (i.e., there are significant changes in the continuum over scales that may not have many high transmission pixels), and our \HI\ Ly$\alpha$ forest continuum fitting algorithm was unable to achieve the same residual level on the mocks, weakening our constraints.\footnote{We altered this algorithm from that described in Section~2.1 to place a continuum point every $3000~$km~s$^{-1}$ rather than $3500~$km~s$^{-1}$ as this provided more degrees of freedom that helped to fit HS1700+6416's more structured continuum.}  Fig.~\ref{fig:eta_measurementHS1700} is similar to Fig.~\ref{fig:eta_measurement} in the main body but instead for HS1700+6416.   The grey vertical bands demarcate regions that have significant features in the continuum.  In the top panel, the thick solid black band is the G140L HS1700+6416 measurement, and the blue solid curve is the \HI--extrapolated \HeII\ absorption for $\eta = 100$, using the analysis pipeline described in the main body.  The dashed blue curves are the same but assume that the continuum fit errs at $\pm2\%$.   In the bottom panel, the black points are $\widehat{\eta}$, the green highlighted region is the statistical error, and the cyan regions again account for a continuum error of $2\%$.  Two per cent is the RMS continuum error in the HS1700+6416 mocks, excluding the grey regions and using a hand-drawn cubic spline fit to the HS1700+6416 continuum as the mocks' underlying continuum.   While the errors are larger than in the measurement towards HE2347-4342, $\widehat{\eta}$ towards HS1700+6416 is again consistent with $\eta = 100$ in most pixels.
 
  \citet{syphers13} reported two quasars that fall $26$ and $20~$comoving Mpc from the HS1700+6416 sightline with estimated redshifts of $2.588$ and $2.625$, respectively (and magnitudes r= 21.2 and 20.9).  Unfortunately, these redshifts fall near the continuum regions that we find most difficult to fit.  Nevertheless, it is worth considering their potential impact on $\eta$.   Using Eq.~(\ref{eqn:proximity}) and simplistically taking the specific intensity to have index $\alpha = 1.6$ to extrapolate from the r band to $4~$Ry, we find that these quasars would have contributed an additional $\Gamma_{\rm HeII}^{\rm quasar} = 0.5\times 10^{-15}~$s$^{-1}$ and $1\times 10^{-15}~$s$^{-1}$ to the sightline at their respective redshifts.  The latter value leads to as much as a factor of two enhancement for our estimate in the text of $\widehat{\Gamma}_{\rm HeII} = (2-4)\times 10^{-15}~$s$^{-1}$.  Intriguingly, there are ${\cal O} (1)$ decreases in $\eta$ centered at very similar redshifts to these quasars, $z=2.58$ and $2.63$ (Fig.~\ref{fig:eta_measurementHS1700}, bottom panel).  These features are not significantly offset from the quasar redshift as we found for the putative proximity zone at $z=2.66$ towards HE2347-4342, and if sourced by these quasars require lifetimes of $>20~$Myr.

\section{B.  Absorber Model}
\label{ap:absorber_model}

Section \ref{sec:model} employed a simple model for the inhomogeneous photoelectric opacity of the IGM that took all absorption systems to have the same power-law density profile of $\rho = \Delta_{0} (r/r_0)^{-\alpha}$.  
 Despite its simplicity, this model can be tuned to match many properties of extragalactic absorptions systems (and those `observed' in simulations; \citealt{furlanetto05, mcquinn-LL}).  Here we elaborate on the details of this model.
 
This model is specified by the power-law index of the \HI\ column density distribution -- $\beta$ --, the mean free path of $1~$or $4~$Ry photons -- $\lambda_{\rm HI}$ or $\lambda_{\rm HeII}$ --, and lastly the density (in units of the cosmic mean) and size of absorbers -- $\Delta_{0}$ and $r_0$.  We do not have complete freedom with any of these numbers as they are constrained by observations.  We take $\beta = 1.7$, consistent with observations \citep{prochaska10, rudie12} and what is found in simulations for systems that cannot self-shield in \HI\ \citep{mcquinn-LL,altay11}.  The results are negligibly affected if instead we take $\beta = 1.5$.  In our model, the power law index of the \HI\ column density distribution, $\beta$, is related to the density profile index, $\alpha$, via  
\begin{equation}
{\alpha} = \frac{1+\beta}{2\,(\beta-1)}.
\end{equation}
This relation can be derived from the differential equation $f(N_{\rm HI}) \, dN_{\rm HI} = 2 \pi r dr$ and the relations $r \propto \rho^{-1/\alpha}$, $N_{\rm HI} \propto \rho^2\, r$.  

The mean free path determines the product of the number density times the size of systems.  To break this degeneracy without adding an additional parameter, we use the model of \citet{schaye01} that posits that the size of absorption systems is the Jeans' length.   This assumption then implies
\begin{eqnarray}
r_0 & = & 0.04  \left(\frac{1+z}{4} \right) \, \left(\frac{ N_{\rm HI}}{10^{17} {\rm \; cm^{-2}}}  \,\frac{\Gamma_{\rm HI}}{10^{-12} {\rm \; s^{-1}}}\right)^{-1/3} \,  \left(\frac{T}{10^4} \right)^{0.41} {\rm ~~comoving~Mpc},\label{eqn:schayer0}\\
\Delta_{0} &= & 200  \left(\frac{1+z}{4} \right)^{-3} \,   \left(\frac{ N_{\rm HI}}{10^{17} {\rm \; cm^{-2}}}  \,\frac{\Gamma_{\rm HI}}{10^{-12} {\rm \; s^{-1}}}\right)^{2/3} \,  \left(\frac{T}{10^4} \right)^{0.17}.\label{eqn:schayerho}
\end{eqnarray}
This ansatz has been found to roughly reproduce the mean properties of absorbers in simulations \citep{mcquinn-LL,altay11}.
In our calculations, $\Delta_{0}$ and $r_0$ are both evaluated at $N_{\rm HI} = 10^{17}~{\rm cm}^{-2}$ and $T = 10^4~$K.  However, for our choice of $\beta = 1.7$, the relations given in equations~(\ref{eqn:schayer0}) and (\ref{eqn:schayerho}) are approximately maintained across the $N_{\rm HI}$ range relevant to the continuum opacity (as consistency would require).   Equations (\ref{eqn:schayer0}) and (\ref{eqn:schayerho}) depend on $\Gamma_{\rm HI}$.  This quantity can be estimated from the emissivity of the sources and the mean free path as $\Gamma_{\rm X} = \int_1^\infty d\nu \, \sigma_{\rm X} \, \nu^{-3} \, \epsilon_\nu \, \lambda_\nu$, where $\sigma_X$ is the photoelectric opacity of species $X$ at its Lyman-limit and $\epsilon_\nu$ is the sources' specific emissivity.  


 This model is now completely specified by the choice of $\lambda_{\rm HI}$ or $\lambda_{\rm HeII}$ and by a model for the sources.  Once these are specified, we populate a $\sim 1$Gpc simulation box with randomly--placed absorbers (with the specified profile and number density) and with sources.  {\comment This randomization is justified in the main body.}   Then, to generate the ionizing background field, we calculate the attenuation along rays from the sources to specified locations in the box.  The results of these calculations are presented in Section~3.

For computational convenience our model makes two additional simplifications.  First, our model calculates the \HI\ or \HeII\ profile of absorbers with the optically thin photoionization rates, which holds at $N_{\rm X} \lesssim \sigma_{\rm X}^{-1}$.  However, we find that even if we aggressively assume that the optical depth goes to $\infty$ when the optically thin 
$N_{\rm X}$ exceeds $\sigma_X^{-1}$, the fluctuations in the ionizing background are negligibly altered.  This finding results because the affected columns already have optical depths $>1$ for photons at the Lyman-limit and so the added opacity has a small impact on their effective cross section.  Second, our absorber models do not account for the response of the absorber sizes to the local ionizing background, instead using the global mean to specify their properties.  
 
\begin{figure}
\begin{center}
{\epsfig{file=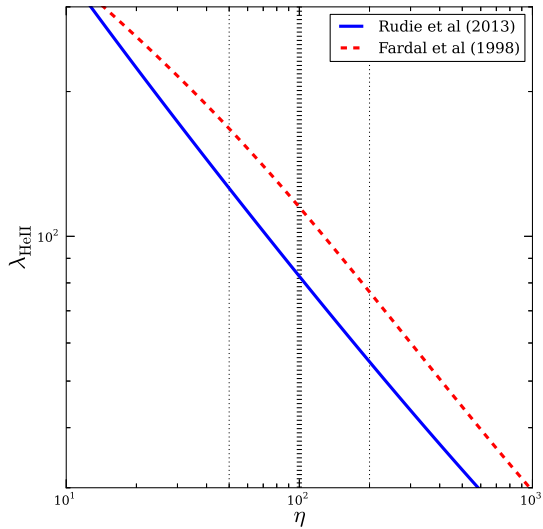, width=8cm}}
\end{center}
\caption{Estimates for the $4~$Ry mean free path as a function of $\eta$ at $z=2.5$, using the column density distribution of \citet[assuming a single power-law for $\beta(N_{\rm HI})$]{rudie13} or  \citet[their A1 model]{fardal98}.  Both curves are calculated for $\Gamma_{\rm HI} = 10^{-12}~$s$^{-1}$ with the algorithm described in Appendix A of \citet{mcquinn-HeII}. \label{fig:mfp}}
\end{figure}

We had mentioned that our absorber model is completely specified for quasar sources by $\lambda_{\rm HeII}$.  Figure~\ref{fig:mfp} shows estimates for $\lambda_{\rm HeII}$ as a function of $\eta$ at $z=2.5$.  These estimates motivate the choice used in much of the paper of $100$~comoving Mpc for $\eta = 100$.  The two curves use different estimates for the column density distribution and calculated using the method first used in \citet{fardal98} and described in Appendix A of \citet{mcquinn-HeII}.  The value of $100$~comoving Mpc is on the small side of recent estimates, agreeing with \citealt{davies12} and a factor of $2$ lower than \citealt{faucher09}.   The difference with \citet{faucher09} primarily owes to the higher value of $\eta$ that we use (and measure).   
 
{\comment For the application to the $4~$Ry background considered in the main body, we fix the abundance of systems with the parameters $\Gamma_{\rm HI} =10^{-12}~$s$^{-1}$ and the $1~$Ry mean free path measurements of \citet{prochaska09}.  We then adjust each absorber's cross section at $4~$Ry to match the desired $\lambda_{\rm HeII}$.}

\section{C: a comment on the impact of small-scale intensity fluctuations in the \HI\ Ly$\alpha$ forest}    
\label{sec:Lyalpha}

We found in Section~\ref{sec:model} that the small-scale intensity fluctuations that owed to absorber discreteness only lead to tens of percent fluctuations in $\widehat{\eta}$; the fluctuations that are most apparent in Fig.~\ref{fig:etaJmocks} owe to quasar proximity regions.  For high--resolution observations of the \HI\ Ly$\alpha$ forest, the small-scale fluctuations in $\Gamma_{\rm HI}$ from absorber discreteness may be a more important consideration, as percent level effects can impact cosmological parameter determinations.  Several studies have investigated the impact of intensity fluctuations on the \HI\ Ly$\alpha$ forest induced by discrete sources \citep{meiksin04, mcdonald05, croft04, mcquinn-Tfluc}, but none has investigated the fluctuations that arise from inhomogeneous attenuation.  We have generated \HI\ Ly$\alpha$ forest mocks with an inhomogeneous $\Gamma_{\rm HI}$.  The $\Gamma_{\rm HI}$ field is calculated using the model presented in Section~\ref{ap:absorber_model}.  We find that this additional source of fluctuations has a small impact on the power spectrum of the \HI\ Ly$\alpha$ forest -- the statistic of choice for most Ly$\alpha$ forest analyses:  It imparts a few percent increase in power at the thermal broadening turnoff in the power spectrum at $z\sim 3$ and, hence, would only slightly bias estimates for, e.g., the temperature of the IGM, and has an even smaller impact on larger scales.  Short timescale fluctuations in quasar light curves ($< 10$~Myr) could potentially have a larger impact on the Ly$\alpha$ forest as these could further enhance the small-scale intensity fluctuations. (The photoionization timescale is $2.5 \, \eta \times$ smaller for \HI\ than \HeII\ making the \HI\ sensitive to shorter timescale variations than the \HeII.)  We refrain from investigating their impact here as it is unclear how to model quasar variability on $\sim 10^4~$yr timescales.

\end{document}